\documentclass[aps,prx,nofootinbib, reprint,superscriptaddress,floatfix,longbibliography]{revtex4-1}
\usepackage{graphicx}
\usepackage{subfigure}
\usepackage{nicefrac}
\usepackage{amsmath}
\usepackage{amssymb}
\usepackage{amsfonts}
\usepackage{amsthm}
\usepackage{mathrsfs}
\usepackage{wasysym}
\usepackage{bm}
\usepackage{placeins}
\usepackage[bookmarks=false]{hyperref}
\usepackage[usenames,dvipsnames]{xcolor}
\usepackage[italicdiff]{physics}
\usepackage{chemformula}
\usepackage{mathtools}
\usepackage{enumitem}
\usepackage{stackengine}
\usepackage[dvipsnames]{xcolor}
\usepackage{soul}

\newcommand{\hc}{\text{H.c.}}
\newcommand{\dg}{\dagger}
\newcommand{\up}{\uparrow}
\newcommand{\down}{\downarrow}

\makeatletter
\renewenvironment{subarray}[2][c]{%
  \if#1c\vcenter\else\vbox\fi\bgroup
  \Let@ \restore@math@cr \default@tag
  \baselineskip\fontdimen10 \scriptfont\tw@
  \advance\baselineskip\fontdimen12 \scriptfont\tw@
  \lineskip\thr@@\fontdimen8 \scriptfont\thr@@
  \lineskiplimit\lineskip
  \ialign\bgroup\ifx c#2\hfil\fi
    $\m@th\scriptstyle##$\hfil\crcr
}{%
  \crcr\egroup\egroup
}
\makeatother

\begin{document}

\title{Hubbard ladders at small $U$ revisited}

\author{Yuval Gannot}
\email{ygannot@stanford.edu}
\affiliation{Department of Physics, Stanford University, Stanford, California 94305, USA}
\author{Yi-Fan Jiang}
\affiliation{Stanford Institute for Materials and Energy Sciences,SLAC National Accelerator Laboratory and Stanford University, Menlo Park, CA 94025, USA}
\author{Steven A. Kivelson}
\affiliation{Department of Physics, Stanford University, Stanford, California 94305, USA}

\date{\today}

\begin{abstract}
We re-examine the zero temperature phase diagram of the two-leg Hubbard ladder in the small $U$ limit, both analytically and using density-matrix renormalization group (DMRG). We find a ubiquitous Luther-Emery phase, but with a crossover in behavior at a characteristic interaction strength, $U^\star$; for $U \gtrsim U^\star$, there is a single emergent correlation length $\log[\xi] \sim 1/U$, characterizing the gapped modes of the system, but for $U\lesssim U^\star$ there is a hierarchy of length scales, differing parametrically in powers of $U$, reflecting a two-step renormalization group flow to the ultimate fixed point. Finally, to illustrate the versatility of the approach developed here, we sketch its implications  for a half-filled triangular lattice Hubbard model on a cylinder, and find results  in conflict with inferences concerning  the small $U$ phase from recent DMRG studies of the same problem.
\end{abstract}

\maketitle

\section{Introduction}
The Hubbard model is of paradigmatic importance in the study of strongly correlated electron systems. From a theoretical standpoint, Hubbard ladders are particularly interesting in that they exhibit aspects of strong coupling physics even at asymptotically weak coupling. From a renormalization group (RG) perspective \cite{Balents_Fisher_1996,Lin_Balents_Fisher_1997,Arrigoni_1996A,Arrigoni_1996B,Schulz_1996,Schulz_1997,Orignac_Giamarchi_1997}, this is reflected in the fact that, in the vicinity of the non-interacting fixed point, the beta function links the flows of multiple coupling constants. In recent years, Hubbard ladders have been the subject of a number of density matrix renormalization group (DMRG) studies at intermediate to strong coupling \cite{Noack1994,Noack_White_Scalapino_1996,Dolfi_Bauer_Keller_Troyer_2015,Liu2012,LeBlanc2015,Ehlers2017,Huang2018,Zheng2017,Jiang2018,Jiang2019,Mishmash_et_al_2015,Szasz_et_al_2020,Shirakawa_et_al_2017,Venderley_Kim_2019,White1997,Dodaro2017,Jiang2017,Jiang2018tj}. In this context, the weak-coupling RG method acquires a newfound significance: assuming adiabatic continuity, it can be used to pin down the phase of matter at the small $U$ end of the range of $U$ that DMRG can accommodate, providing an important point of reference for DMRG calculations.

With 
these motivations in mind, we study the small-$U$ limit of the $N$-leg Hubbard ladder in two special instances. First, we revisit the two-leg ladder using both weak-coupling RG and DMRG. Having developed a general understanding from the in-depth study of 
this example problem, we then consider (in somewhat less detail) the triangular lattice Hubbard model on a four-leg cylinder, which has been the subject of recent DMRG studies \cite{Szasz_et_al_2020,Shirakawa_et_al_2017,Venderley_Kim_2019}.

The RG approach to the study of multi-leg Hubbard ladders was pioneered by Balents and Fisher (BF) in their analysis of the two-leg ladder \cite{Balents_Fisher_1996}, and later extended to ladders with arbitrary numbers of legs by Lin, Balents, and Fisher \cite{Lin_Balents_Fisher_1997}. Using the one-loop beta function, which is valid for arbitrarily weak interactions, BF determined which of the interactions grow most strongly in the course of the RG flow away from the non-interacting fixed point. By examining the nature of these most rapidly growing or ``dominant'' interactions, they were able to map out 
a conjectural ground state phase diagram of the system.

Interestingly, BF noticed that very different results are obtained 
depending on whether one starts with initial interactions that are asymptotically small ($U \lesssim U^\star \sim 10^{-5}$ in units in which the rung-hopping matrix element is $t=1$) or only pretty small ($U^\star \lesssim U \ll 1$). The result in the latter case implies a so-called C1S0 (Luther-Emery liquid \cite{Luther_Emery_1974}) phase with a single gapless charge mode and a spin gap, the one-dimensional analogue of a superconductor. On the other hand, when $U \lesssim U^{\star}$ the dominant interactions suggest a so-called C2S1 state with two gapless charge modes and one gapless spin mode. BF therefore conjectured that in the true weak-coupling limit, the C2S1 phase is the 
ground state of the system.

\begin{figure}
\centering
\includegraphics[width=\linewidth]{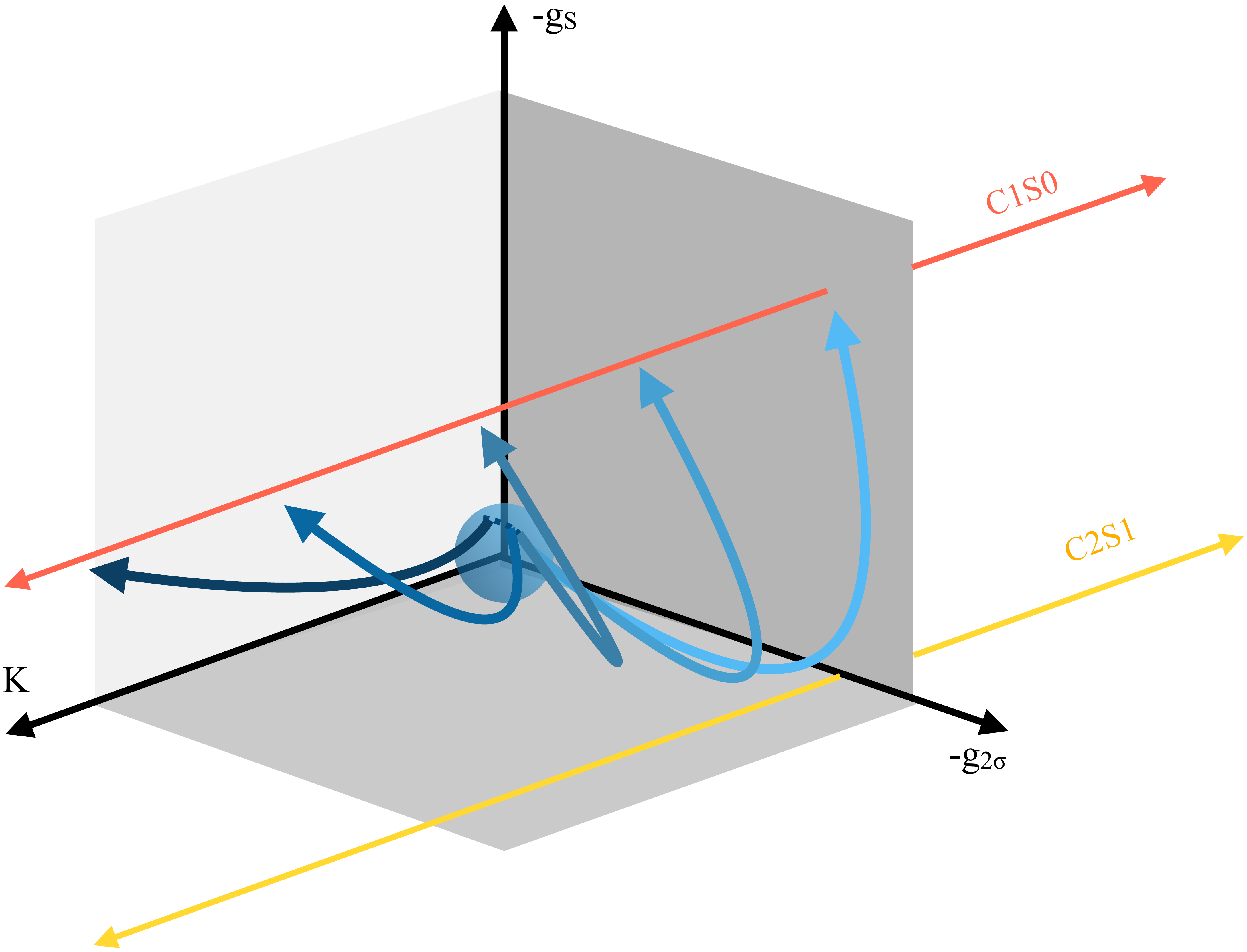}
\caption{Schematic global flow diagram.  The dimensionless running coupling constants $g_{2\sigma}$ and $g_S$ are defined in Sec. \ref{sec:Hamiltonian}. The blue lines indicate RG flows for different initial conditions 
with the darker blue corresponding to larger values of $U/U^{\star}$.
Here the lines labeled  C2S1 and C1S0 represent the fixed ``lines'' discussed in the main text.
The sphere surrounding the non-interacting unstable fixed point at $U=0$ indicates the regime in which we explicitly follow these flows using the leading order perturbative expression for the beta function.  As indicated, even when the initial flows approach the C2S1 fixed line, they do so in a regime in which it is perturbatively unstable, implying that the ultimate long-distance behavior is controlled by the C1S0 fixed line. Different points on the C1S0 fixed line correspond to different values of the Luttinger exponent, $K$, defined in Eq. \ref{2K}.
\label{fig:flowdiagram}}
\end{figure}

In the present paper, we extend and correct this important work. We begin by re-analyzing the RG flows using a recently developed \cite{Vafek_Yang_2010,Vafek_2010} improved method for analyzing the sort of RG flows that arise for multiple intertwined interactions, clarifying the origin of the unexpected result that the flow away from the non-interacting fixed point depends on the strength of the initial interactions.  
We then reexamine the BF conjecture for the C2S1 ground-state in the true weak-coupling limit, finding that the interplay between the different ordering tendencies is more subtle. In common with other gapless critical phases in one-dimension, the C2S1 phase is associated not with an isolated RG fixed point but instead with a multi-dimensional surface of fixed points, parameterized by marginal operators (analogues of the Luttinger parameter) which determine the local stability of the surface. Regardless of its dimension, we from now on refer to a surface of this sort as a ``fixed line''.
For $U \lesssim U^{\star}$ the flow away from the non-interacting fixed point is toward a point on the C2S1 fixed line which is itself perturbatively unstable with respect to one of the subdominant interactions. This suggests, as first discussed in Ref. \onlinecite{Emery_Kivelson_Zachar_1999}, that a second stage RG flow carries the system away from the C2S1 line and toward the C1S0 line.
This is illustrated by the schematic global RG flow in Fig.~\ref{fig:flowdiagram}. 

The upshot of this analysis is that the C2S1 phase conjectured at $U \lesssim U^{\star}$ is replaced by the C1S0 phase. However, it exhibits a hierarchy of energy scales. In agreement with BF, we find that the dominant gap is exponentially small in $1/U$. However, we find that there are additional parametrically smaller gaps down by factors of $\sqrt{U/U^{\star}}$. These subdominant gaps are associated with operators that are effectively irrelevant in the vicinity of the weak-coupling fixed point, but which become relevant upon approach to the unstable region of the C2S1 fixed line; they 
are associated with the second stage RG flow from the intermediate unstable C2S1 fixed 
line to the stable C1S0 fixed fixed 
line.  In contrast, in the $U \gtrsim U^{\star}$ regime, the gaps associated with different modes are roughly the same order of magnitude in size, reflecting a direct flow to the C1S0 fixed line. 

The physical interpretation of $U^\star$ is therefore not a phase transition, as in the BF conjecture, but a crossover, such that for $U \lesssim U^\star$, there is a large difference in the gap scales (or correlation lengths) of various different correlations, while for $U \gtrsim U^\star$, there is a single emergent length scale that characterizes the fall off of correlations. (Additionally, the superconducting and CDW correlations exhibit power-law fall-off at long distances, reflecting the existence of the gapless charge mode).

For $U \lesssim U^\star$, we demonstrate the existence of the subdominant gaps by explicitly following the perturbative RG flows up to a point which is sufficiently far from the non-interacting fixed point that the distinction between the dominant and subdominant interactions is established, yet close enough that the perturbative approach remains valid. We solve the resulting Hamiltonian in a mean-field approximation. The result passes an important self-consistency check in that the predictions for the gap magnitudes are independent of where we terminate the RG flows. Along the way, we also determine how $U^\star$ varies with the velocities of the two bands, and in particular show that it vanishes exponentially as one approaches a symmetric condition in which the Fermi velocities on the bonding and anti-bonding bands are equal.

Next, we report the results of a DMRG study of the two-leg ladder at $U = 4$. In agreement with the weak-coupling theory and with previous DMRG studies at ${U=8}$ \cite{Noack_White_Scalapino_1996,Dolfi_Bauer_Keller_Troyer_2015}, we find a ubiquitous Luther-Emery liquid. Though 
$U=4$ cannot truly  said to be ``weak'',   it is apparently small enough that we see clear vestiges of the non-interacting Fermi surface, 
{\it i.e.} the single-particle occupancy in momentum space, $n(\bm{k})$, exhibits a sharp drop at the non-interacting Fermi momenta~\footnote{This is to be contrasted with the case of much larger $U = 12$, considered in Fig.~\ref{Afig:nkU12} of Appendix \ref{app:numerical_details},
where the features in $n(\bm{k})$ are highly rounded and centered away from the non-interacting Fermi momenta}. Thus, it is reasonable to interpret these results in terms of modes corresponding roughly to particle-hole excitations about the Fermi surface, as is assumed in the weak-coupling RG approach. Nevertheless, by examining very large system sizes (up to $L_x = 288$) we determine from the spin correlations and the central charge that the system is in a C1S0 phase.

Finally, without going into the same level of detailed analysis, we consider the small $U$ limit of the triangular lattice Hubbard model on a four-leg cylinder.  Exciting results concerning the possible existence of a chiral spin liquid phase at intermediate $U$ have been obtained \cite{Szasz_et_al_2020} from recent DMRG studies of this problem 
at an electron density of $n=1$ electron per site.  These same calculations have identified the small $U$ phase of this system as a C3S3 phase with gapless quasi-particle excitations at all the Fermi crossings of the non-interacting problem.  Our analysis shows, instead, that  the small $U$ phase  of this problem is a chiral Luther-Emery liquid (C1S0) which can be visualized  as a fluctuating  $d+id$ superconductor on a finite cylinder.  (Note, it has been previously shown \cite{Raghu_Kivelson_Scalapino_2010, Nandkishore_Thomale_Chubukov_2014} 
that the ground-state of the 2D problem at small $U$ has $d+id$ off-diagonal long-range order.)  We suggest that the putative Fermi surfaces identified in the DMRG study are actually not true singularities, but rather vestigial features similar to what we have seen in the two-leg ladder at $U=4$.  We will report in more detail on the application of the present analysis to this problem in a future publication \cite{Gannot_Jiang_Kivelson_2020}.  Among other things, it is worth noting that a transition in the Kosterlitz-Thouless universality class between a chiral Luther-Emery liquid and a chiral spin-liquid can be accounted for under the supposition that an Umklapp scattering term changes from being irrelevant to relevant at a critical value of $U$.

The remainder of this paper is organized as follows: In Section~\ref{sec:Hamiltonian}, we introduce the Hamiltonian for the two-leg ladder and define the dimensionless coupling constants which enter the weak-coupling continuum limit. In Section~\ref{sec:RG_analysis}, we analyze the formal solutions to the one-loop RG equations of BF using the method of Refs.~\cite{Vafek_2010,Vafek_Yang_2010}. This formal analysis leads to a renormalized Hamiltonian which we study in Section~\ref{sec:analysis}, explaining how the C2S1 phase reduces to a multi-scale C1S0 phase. We discuss also the nature of the CDW and superconducting correlations. The global RG flow shown in Fig.~\ref{fig:flowdiagram} is discussed in Section~\ref{sec:global_RG}.
In Section \ref{sec:DMRG}, we report the results of our DMRG study of the two-leg ladder. Finally, in Section \ref{sec:triangular} we perform a weak coupling analysis of the triangular lattice Hubbard cylinder and discuss its relation to recent DMRG studies.

\section{Two leg ladder Hamiltonian and continuum limit
\label{sec:Hamiltonian}
}
The Hamiltonian for the two-leg Hubbard ladder with repulsive interactions is $H = H_0 + H_{\text{int}}$, where
\begin{align}
 &H_0 =\begin{multlined}[t]
    -\sum_{x,j,\alpha} \pqty{c^{\dg}_{x+1,j,\alpha}c_{x,j,\alpha} + \hc}\\
    -t_{\perp}\sum_{x\alpha} \pqty{c^{\dg}_{x,2,\alpha}c_{x,1,\alpha}+ \hc},
\end{multlined} \\
    &H_{\text{int}} = U\sum_{xj} :(c^{\dg}_{x,j,\up}c_{x,j,\up})
    (c^{\dg}_{x,j,\down}c_{x,j,\down}):,
    \label{eq:Hamiltonian}
\end{align}
and $U>0$. Above, $c_{x,j,\alpha }$ annihilates an electron with spin $\alpha = \up, \down$ on leg $j=1,2$ at position $x$ along the chain, and $:\hspace{4 pt}:$ denotes normal ordering. 

For small $U$, it is appropriate to first diagonalize the non-interacting piece of the Hamiltonian. Thus we introduce  anti-bonding ($i = 1$) and bonding ($i=2$) orbitals
\begin{equation}
    \phi_{x,i,\alpha} = \frac{1}{\sqrt{2}}\pqty{c_{x,1, \alpha }+(-1)^{i}c_{x,2, \alpha }}. \label{eq:orbitals}
\end{equation}
which,  when Fourier transformed in the $x$-direction, yield the band energies as a function of the Bloch wave-vector $-\pi < k \leq \pi$:
\begin{equation}
    \epsilon_i(k)  = (-1)^{i}t_{\perp} -2\cos(k).
\end{equation}
Next, we focus our attention on the low energy degrees of freedom near the Fermi points, introducing continuum left and right movers $\psi_{Li\alpha}$ and $\psi_{Ri\alpha}$ as
\begin{equation}
    \phi_{i \alpha x} \approx \psi_{R i \alpha}(x)e^{i k_{Fi} x} + \psi_{L i \alpha}(x)e^{-i k_{Fi} x},
\end{equation}
where $k_{Fi}$ is the Fermi momentum of band $i$. This gives
\begin{equation}
H_0 \approx \sum_{i,\alpha} \int dx \hspace{2 pt} v_{i}(\psi^{\dg}_{L i\alpha} i \partial_{x} \psi_{L i \alpha} - \psi^{\dg}_{R i\alpha} i \partial_{x} \psi_{R i \alpha}),
\end{equation}
where $v_i = 2\sin{k_{Fi}}$ is the Fermi velocity, and from now on we consider only the 
range of electron densities per site, $n$, where both bands are partially filled.

The interaction density is conveniently expressed as
\begin{widetext}
\begin{align}
-\mathcal{H}_{\text{int}} &= \widetilde{g}_{1\rho} J_{1R}J_{1L} + \widetilde{g}_{2\rho} J_{2R}J_{2L} + \widetilde{g}_{x\rho} (J_{1R}J_{2L} + J_{2R}J_{1L})\nonumber \\
&+ \widetilde{g}_{1\sigma} \bm{J}_{1R} \vdot  \bm{J}_{1L} + \widetilde{g}_{2\sigma} \bm{J}_{2R} \vdot  \bm{J}_{2L} \nonumber + \widetilde{g}_{x\sigma} (\bm{J}_{1R} \vdot  \bm{J}_{2L} + \bm{J}_{2R} \vdot  \bm{J}_{1L} ) \\
&+ \widetilde{g}_{S}(O_{2S}^{\dg}O_{1S} +\hc)
+ \widetilde{g}_{T}(\bm{O}_{2T}^{\dg} \vdot \bm{O}_{1T} +\hc), \label{eq:interaction_density}
\end{align}
\end{widetext}
where
\begin{align}
    J_{i R}&= :\psi^{\dg}_{R i\alpha} \psi_{R i \alpha}:, \\
    \bm{J}_{ i R} &= :\psi^{\dg}_{R i\alpha}\frac{\bm{\sigma}_{\alpha\beta}}{2} \psi_{R i \beta}:,
\end{align}
(and the same for $R \leftrightarrow L$) are,
respectively, the
spin and charge currents, and
\begin{align}
    O_{iS} &= \frac{1}{\sqrt{2}}\psi_{Ri\alpha}\varepsilon_{\alpha\beta}\psi_{Li\alpha} \\
    \label{SCOP}
    \bm{O}_{iT} &= \frac{1}{\sqrt{2}}\psi_{Ri\alpha} (\bm{\sigma}\varepsilon)_{\alpha\beta}\psi_{Li\alpha},
\end{align}
are singlet and triplet pairing operators. For generic $n$, the Fermi momenta do not satisfy any commensurability relations, and thus there are no Umklapp processes.  Up to purely chiral interactions which may be neglected in a lowest order treatment, Eq.~(\ref{eq:interaction_density}) is then the most general non-irrelevant interaction allowed by spin and crystal momentum conservation.  For the Hubbard model, the  
bare couplings are $\widetilde{g}_{i\rho} = \widetilde{g}_{x\rho} = -U/4$, $\widetilde{g}_{i\sigma} = \widetilde{g}_{x\sigma} = U$, $\widetilde{g}_{S} = -U$, and $\widetilde{g}_{T}= 0$. We also define for convenience below the dimensionless couplings
\begin{equation}
    g_a = \begin{cases}
    {\widetilde{g}_{a}}/(2\pi v_i) & \text{if $a=i\rho,i\sigma$} \\
    {\widetilde{g}_{a}}/(\pi( v_1+v_2)) & \text{else}.
    \end{cases}
\end{equation}
A comparison of our conventions with those of BF may be found in Appendix \ref{app:RG}.

\section{Analysis of the RG equations \label{sec:RG_analysis}}

Following BF, we use the perturbative RG to track the evolution of the coupling constants under an increase in length scale. In common with other problems with multiple naively marginal interactions, the general form of the RG equations is
\begin{equation}
    \dv{{g}_{a}}{\ell} = A_{a}^{bc}g_b g_c + \ldots .
    \label{eq:general_one_loop}
\end{equation}
where $\dd \ell$ is the fractional increase in the length scale being probed, the tensor $A$ encodes the results of a leading order (one-loop) perturbative analysis and $\ldots$ signifies higher order terms in powers of $g_a$ (which we will ignore). The one-loop RG equations for the two-leg ladder were derived in detail by BF, and are presented explicitly in Appendix~\ref{app:RG}.

\subsection{Strategy}

Depending on the initial conditions, the solutions to an equation of the form (\ref{eq:general_one_loop}) may diverge at some finite $\ell_{\infty}$. In the present context, this would signal an instability of the non-interacting fixed point. Of course, the RG equations are valid only as long as the renormalized couplings are small. Therefore any formal, diverging solution $g_a(\ell)$ is meaningful only while 
\begin{equation}
    \max\{ g_a(\ell) \} \ll 1 \label{eq:perturbative_validity}
\end{equation}
However, as $U \to 0$ one can probe arbitrarily near the divergence point before the perturbative approach starts to break down. Hence, in the true weak-coupling limit what matters is the asymptotic behavior of the formal solutions.

BF introduced an elegant method for analyzing this asymptotic behavior. They noted that there is an important class of exact solutions, the so-called rays:
\begin{equation}
    g_a = \frac{G_a}{\ell_{\infty}-\ell}.
    \label{eq:ray}
\end{equation}
Plugging this ansatz into the general form~(\ref{eq:general_one_loop}), the allowed rays correspond to solutions of $G_a= A_a^{bc}G_b G_c$, of which there are finitely many. Asymptotically, a diverging solution always renormalizes onto one of these rays, in the sense that
\begin{equation}
\underset{\ell \to \ell_{\infty}} \lim (\ell_{\infty}-\ell)g_a = G_a.
\end{equation}
That is, 
the interactions grow in a ``direction'' in interaction space that  
is increasingly parallel to one of the rays.  We will call $g_a$  a subdominant coupling if it vanishes exactly on the asymptotic ray, i.e. if $G_a = 0$.
It follows that in the weak coupling limit, the couplings can flow out from  the non-interacting fixed point along only finitely many possible directions. Which one is picked out depends on the initial conditions. The phase of matter exactly on one of these rays is typically straightforward to determine.

The rest of this section is organized as follows. Below, we review 
the ray solutions that appear 
in the two-leg ladder. Then, we use the method of Vafek and Yang~\cite{Vafek_Yang_2010} to clarify what happens when $U$ is small but finite. Finally,
we study the 
evolution of the subdominant couplings, associated with  deviations from a given ray.

\subsection{Rays for the two-leg ladder}
Taking for initial conditions the bare couplings of the Hubbard model, the solutions in which we are interested form a two-parameter family which can be labeled by the velocity ratio \begin{equation}
     r = v_2/v_1
 \end{equation}
 and 
 \begin{equation}
    u = U/(\pi(v_1+v_2)).
\end{equation}
The beta function depends only on $r$; the initial conditions depend on both $u$ (which fixes their magnitude) and $r$ (which fixes their direction). We will denote these solutions as $g_a(\ell;u)$, suppressing the parametric dependence on $r$. We now review the asymptotic behavior of these solutions, as revealed by a direct numerical integration of the RG equations. 
From now on we shall consider only $r\ge 1$, as
 results for $r \to 1/r$ can be obtained by swapping the two band indices.

For $r \gtrsim 8.6$ (i.e. when the Fermi energy is close 
to the edge of band~$1$), the couplings do not flow out along any ray; instead they remain $\order{u}$ and approach a weak coupling fixed line with marginal interactions with strength proportional to $u$. Thus, the RG treatment is perturbatively controlled. Moreover, at the fixed line only $g_{i\rho}$ and $g_{x\rho}$ are nonzero. BF showed that this corresponds to a C2S2 (generalized Luttinger Liquid) phase.

For $1< r \lesssim 8.6$, on the other hand, the solution always diverges at finite $\ell=\ell_{\infty}$. In this range there are two important rays, which we denote  by C2S1 and C1S0 according to the strong coupling fixed 
lines they point towards.  On the C2S1 ray,  $G_{2\sigma} = -1$ and the remaining $G_a$ are zero. On the C1S0 ray, $G_{x \sigma} = G_{T} = 0$ and the remaining $G_a$ are nonzero 
and vary continuously with~$r$. Their explicit expressions can be found in Appendix \ref{app:RG_details}, but for 
present purposes we will need only their signs, which are listed in  Table \ref{tab:rays}.

\begin{table}[]
\setlength\tabcolsep{3pt}
\begin{tabular}{|l|llllllll|} 
    \hline 
    Ray & $G_{1\rho}$ & $G_{2\rho}$ & $G_{x\rho}$ & $G_{1\sigma}$ & $G_{2\sigma}$ & $G_{x\sigma}$ & $G_{S}$ & $G_{T}$ \\
\hline
C2S1 & $0$ & $0$  &  $0$  & $0$  & $-$ & $0$  & $0$ & $0$  \\
C1S0 & $+$ & $+$ & $-$ & $-$ & $-$  & $0$ & $-$ & $0$\\
    \hline 
\end{tabular}
\caption{\label{tab:rays}
Signs of the nonzero $G_a$ for the C2S1 and C1S0 rays. For the C2S1 ray, we always have $G_{2\sigma} = -1$. For the C1S0 ray, these are functions of $r$, which may be found in Appendix \ref{app:RG} but which will not be needed for present purposes.}
\end{table}

Exactly at $r=1$ the asymptotic ray is  C1S0, while for $r$ near $8.6$ it is C2S1. What happens in between is harder to discern.
Lin, Balents, and Fisher showed analytically that for any $r>1$ the C1S0 ray is asymptotically unstable, giving way to the C2S1 ray. However, this crossover is difficult to study numerically, because it is pushed out to the divergence point as $r \to 1$. 

Below, we achieve a 
 clearer window into the asymptotic regime using a different approach, first used by Vafek and Yang \cite{Vafek_Yang_2010} to study an RG equation of the form (\ref{eq:general_one_loop}) in the context of a 2D quadratic band crossing.

\subsection{New approach \label{sec:new_analysis}}
The idea  of Vafek and Yang is to
express the RG flows as a function of the most divergent coupling constant rather than of the scaling parameter, $\ell$. In the present case this is $g_{2\sigma}$ which, over the entire range ${1 \le r \lesssim 8.6}$, diverges to $-\infty$. As shown in
Ref.~\onlinecite{Vafek_Yang_2010}, the ratios $f_a = g_a/g_{2\sigma}$ satisfy a flow equation of the form
\begin{equation}
     \dv{f_a}{x} 
    = F_a(\{f_b\}), \label{eq:ratio_RG}
\end{equation} 
where
\begin{subequations}
\begin{align}
x &= \log(\abs{g_{2\sigma}/g_{2\sigma}(0;u)})  \\
&= \log( \abs{g_{2\sigma}}(2\pi v_2/U))
\label{eq:x_def}
\end{align}
\end{subequations}
and  importantly $F_a$ does not explicitly depend on $x$. 
Notice that
a ray solution to the original RG  equations 
corresponds to a 
fixed value of the ratios $f_a = G_a/G_{2\sigma}$.

A technical difficulty is that this change of variables is only possible so long as $g_{2\sigma}$ is a monotonic function of $\ell$ which does not cross zero.  In the present case, for initial conditions corresponding to the Hubbard model $g_{2\sigma}$  starts off positive and then crosses zero. Thus, before transforming to the new variables, we need determine the initial flow by  integrating the original equations up to a point at which $g_{2\sigma} < 0$. 
  To be explicit, we integrate the original RG equations up to the  scale $\ell_0(u)$ at which $g_{2\sigma}$ is negative but with magnitude equal to its starting value:
\begin{equation}
g_{2\sigma}\pqty{\ell_0(u);u} = -g_{2\sigma}\pqty{0;u}.
\end{equation}
We then use the values of $f_a=g_a\pqty{\ell_0(u);u}/g_{2\sigma}\pqty{\ell_0(u);u}$ as initial conditions for computing the remaining flows from Eq. \ref{eq:ratio_RG}.

\begin{figure}
\centering
\includegraphics[width=\linewidth ]{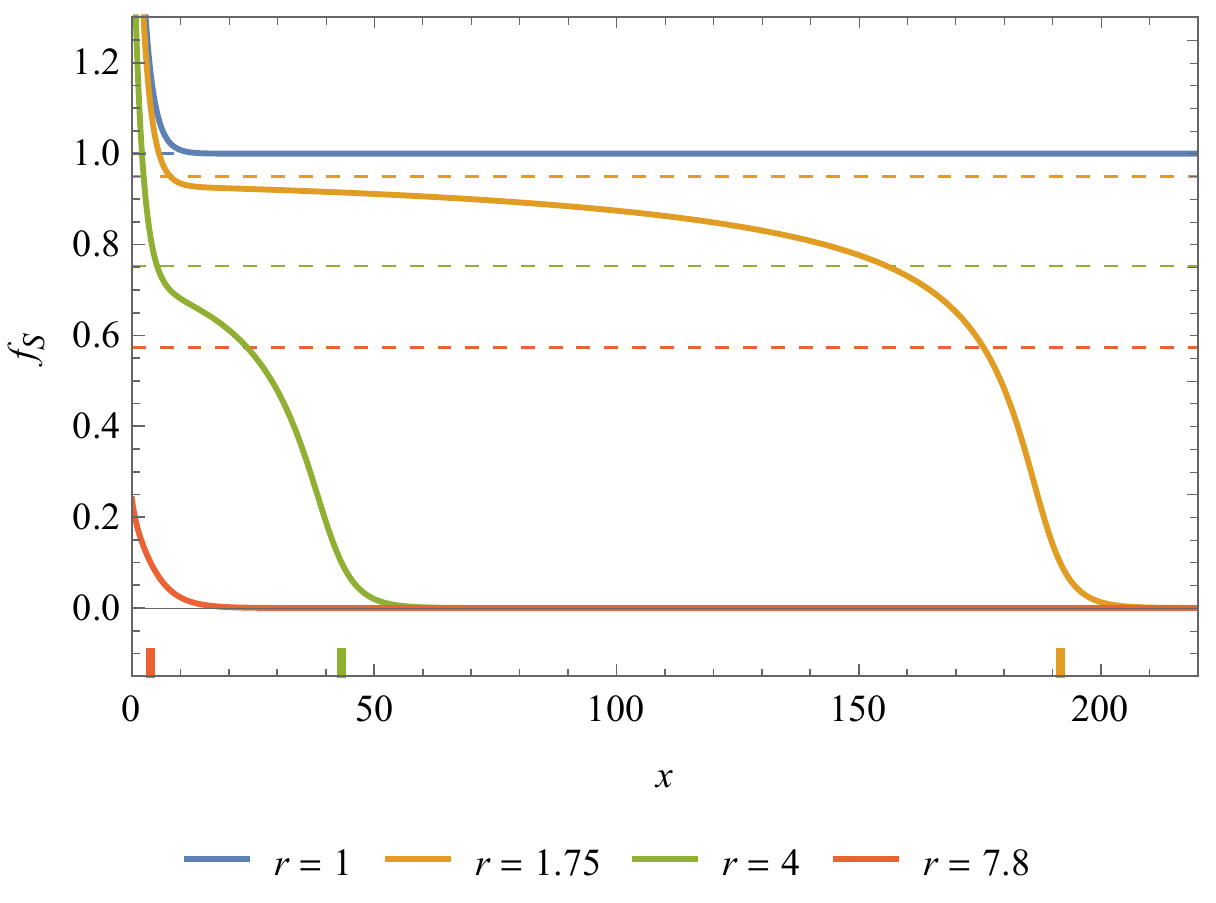}
\caption{Plots of $f_S(x)$ for different values of $r$, represented by different colors. For each color/value of $r$, the dashed line denotes the value on the C1S0 ray and the tick-mark on the $x$-scale denotes $x^{\star}$, defined in Eq.~(\ref{eq:x_star_def}). The C2S1 ray is at zero.
\label{fig:ratios}}
\end{figure}

The resulting flows of the various ratios $f_a$ are then readily computed, with the results shown in Appendix~\ref{app:RG_details}. Here we focus on one of these, $f_S$, since -- as we shall see in the following section -- the singlet pair tunneling interaction $g_S$ is the most important of the subdominant couplings. Specifically, whereas exactly along the C2S1 ray one obtains a C2S1 phase, the coupling $g_S$ is responsible for the instability that ultimately drives the system to a C1S0 phase.
Thus, in Fig.~\ref{fig:ratios} we show $f_S(x)$ for different values of $r$. For $r=1$, $f_S$ asymptotes to the C1S0 ray, whereas for any ${r > 1}$ it eventually decays to zero. However, as $r \to 1$ there is 
a long plateau increasingly near the C1S0 ray. This behavior reflects the known result that the C1S0 ray is stable at $r =1$ but otherwise unstable.

We now discuss how the scale $U^{\star}$ appears. From Eq.~(\ref{eq:x_def}), we see that the smaller the value of $U$, 
the larger the value of $x$ one can reach before $\max \{\abs{g_a} \} = \abs{g_{2\sigma}}$ gets too large and the perturbative approach starts to break down. This reflects the fact that the asymptotic behavior of the formal solutions is, in general, accessible only for arbitrarily small $U$.

Specifically, while the flows are well-defined for all values of $g_a$, the perturbative RG equations they encode are only reliable so long as all $|g_\sigma| \lesssim g$ where $g$ is  small.  Thus, we must  stop this analysis when $|g_{2s}| =g$. If we let $x^{\star}$ denote the characteristic value of $x$ at which point $f_S$ drops
from its plateau value, then there must be a corresponding scale $U^{\star}$ whose meaning is as follows: For $U \lesssim U^{\star}$ the range ${x \gtrsim x^{\star}}$ is accessible, meaning 
that the flows reach the vicinity of the C2S1 ray while the interactions are all still weak.  For $U \gtrsim U^{\star}$, on the other hand, only $x \lesssim x^{\star}$ is accessible 
implying that the flow 
is still close to the C1S0 ray when $|g_{2\sigma}| =g$. The scale $U^{\star}$ will be identified below with the crossover between single-scale and multi-scale C1S0 regimes.

The explicit relationship between $U^{\star}$ and $x^{\star}$ is
\begin{equation}
U^{\star} \sim  2\pi v_2 g e^{-x^{\star}},
\end{equation}
where $g$ is the value of $ \abs{g_{2\sigma}}$ at which point the perturbative RG first starts to break down. The ambiguity in $g$ is unimportant since $x^{\star}$ is also not sharply defined and appears in the exponent. Therefore, to logarithmic accuracy, we may write
\begin{equation}
U^{\star} \sim  v_2 e^{-x^{\star}}. \label{eq:U_star_def}   
\end{equation}

\begin{figure}
\centering
\includegraphics[width= \linewidth]{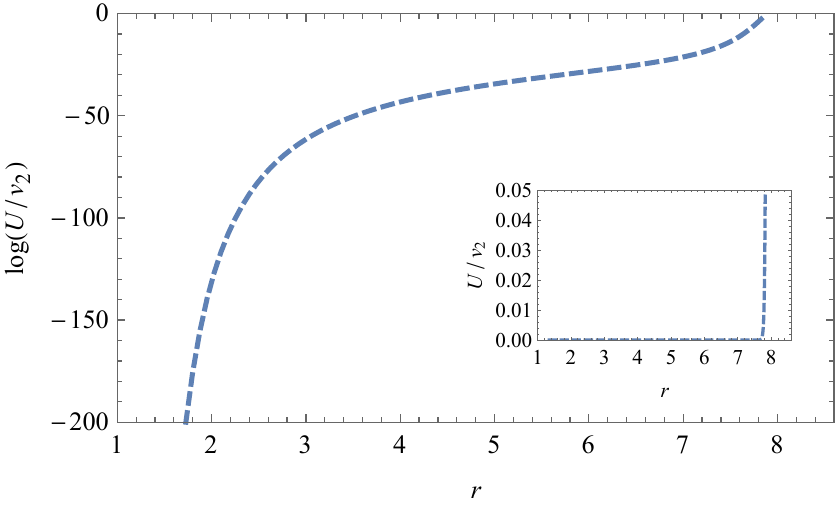}
\caption{
\label{fig:crossover}
Main plot: $\log(U^{\star}/v_2) = -x^{\star}$. Inset: $U^{\star}/v_2 = e^{-x^{\star}}$. Here $x^{\star}$ is defined according to Eq.~(\ref{eq:x_star_def}). The C2S1 ray is perturbatively accessible only below the crossover line.
}
\end{figure}

Next we determine $x^{\star}$ and thus $U^{\star}$ as a function of $r$. To be explicit, we define $x^{\star}$ according to
\begin{equation}
|f_S(x^{\star})| = 0.1 \label{eq:x_star_def}.
\end{equation}
In Fig.~\ref{fig:ratios}, the resulting value of $x^{\star}$ is indicated for each value of $r$ by a tick mark of the appropriate color. Then in Fig.~\ref{fig:crossover}, we plot $-x^{\star} = \log(U^{\star}/v_2)$ in the main panel and $e^{-x^{\star}}=U^{\star}/v_2$ in the inset, both as functions of $r$.

We now summarize the $r$-dependence of $x^{\star}$ and $U^{\star}$ shown in Figs. 2 and 3. Consider first $r \to 1$, where $x^{\star}$ diverges. In Appendix \ref{app:RG_details}, we show that
$
x^{\star}\sim 1/(r-1)^2.
$
Thus $U^{\star}$ vanishes as
\begin{eqnarray}
U^{\star} \sim e^{-c/(r-1)^2},
\end{eqnarray}
for some positive constant $c$. Upon increasing $r$, $x^{\star}$ remains rather large at first (meaning $U^{\star}$ remains very small), but eventually drops to zero at $r \approx 7.8$. At this point $U^{\star}$ bends sharply upward, mimicking a vertical crossover line. 

Once $r \gtrsim 7.8$, we can see from Fig.~\ref{fig:ratios} that the flow away from the non-interacting fixed point is always along the C2S1 ray. In this regime, $U^{\star}$ is not a physically meaningful scale for the small $U$ problem. However, to simplify the notation below we will use  Eq.~(\ref{eq:U_star_def}) and (\ref{eq:x_star_def}) to set $U^{\star} \sim v_2$ in this regime.

The remaining ratios (shown in Appendix~\ref{app:RG_details}) behave similarly to $f_S$; in particular, they all first become small at roughly the same $x^{\star}$. The important exceptions are $f_{x\sigma}$ and $f_{T}$, which in all cases rapidly decay to zero. As noted by BF, this reflects an emergent approximate conservation of spin within each band.

\begin{figure}
\centering
\includegraphics[width=\linewidth]{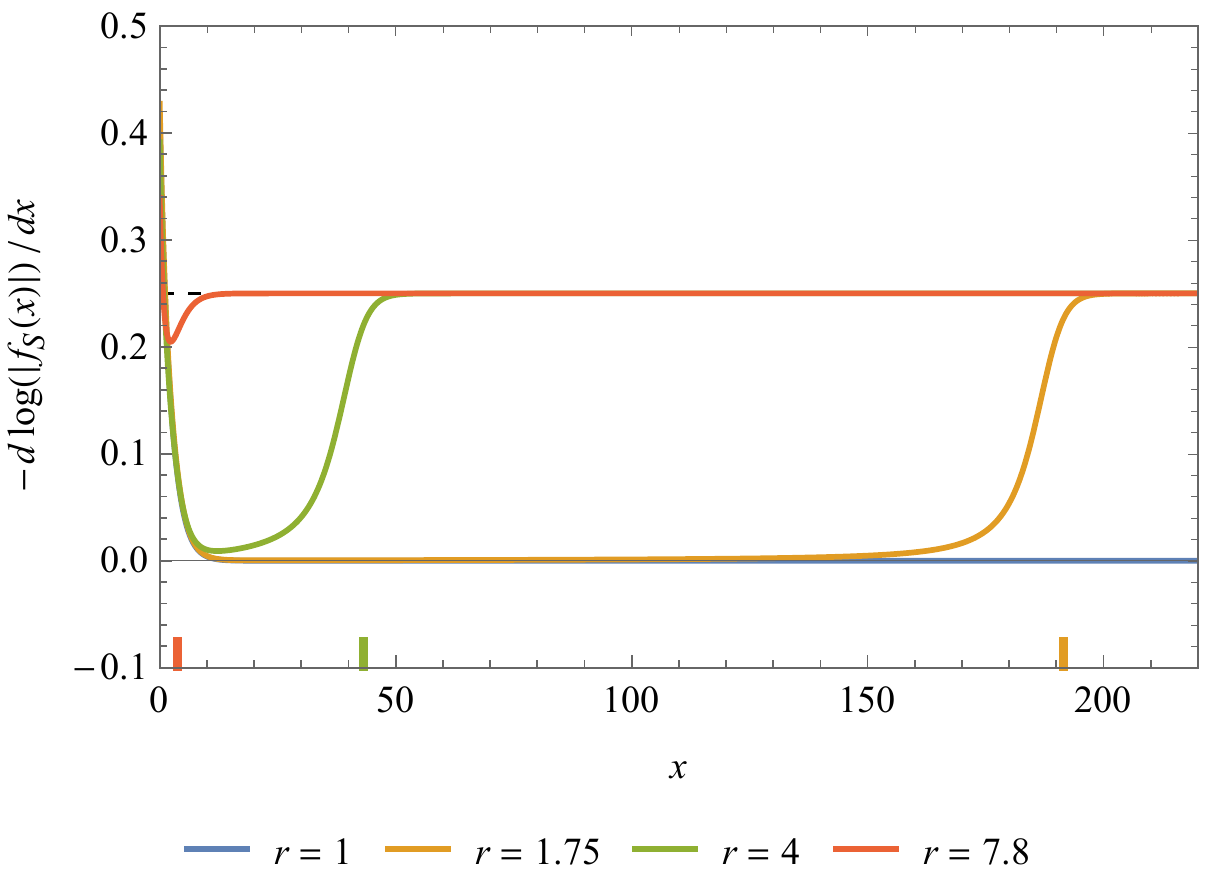}
\caption{Plot of $ -\dv*{\log(|f_S(x)|)}{x}$, for which the limit as $x\to \infty$ is $\lambda_S$. As in Fig.~(\ref{fig:ratios}), the tick marks on the $x$-scale denote $x^{\star}$ as defined in Eq.~(\ref{eq:x_star_def}).
\label{fig:exp}}
\end{figure}

Finally, we determine the subdominant couplings 
as the flows approach the C2S1 ray. 
Near this ray,
the ratios of the subdominant to dominant couplings decay as ${f_a \sim e^{-\lambda_a (x-x^{\star})}}$ for some $\lambda_a > 0$, and therefore
\begin{equation}
g_{a} \sim g_{2\sigma} \pqty{\frac{U}{U^{\star}} \frac{1}{\abs{g_{2\sigma}}}}^{\lambda_a}.
\label{eq:subdominant}
\end{equation}
This form holds when the quantity inside the parentheses is small, for which a necessary condition is 
  $ U \ll U^{\star}$. 
The exponents $\lambda_a$ are the limit as $x \to \infty$ of $ -\dv*{\log(|f_a(x)|)}{x}$. Focusing once more on the singlet pair tunneling term, we plot $ -\dv*{\log(|f_S(x)|)}{x}$
for various $r$ in Fig.~(\ref{fig:exp}). 
 Notice that 
 in the large $x$ limit, 
 the curves all approach
 $\lambda_S=1/4$  independent of $r$. 
The remaining $\lambda_a$ are also $r$-independent quantities. Their values are listed in Appendix~\ref{app:RG_details}, but will not be needed for the present purposes.

\subsection{Summary}
As the analysis discussed in this section is somewhat complicated, we summarize the conclusions. What we have done is to integrate the RG equations for fixed velocity ratio in the range $1 < r \lesssim 8.6$ and for initial conditions corresponding to the Hubbard model with small $U$. Since the RG equations are only valid for $\abs{g_a} \ll 1$, we can follow the flows in this manner only in the vicinity of the non-interacting fixed point.  

To be concrete, we on terminate the flows at the RG scale $\ell = \ell_f$ where the dominant interaction $g_{2\sigma}$ satisfies $g_{2\sigma}(\ell_f) = -g$, where $g$ is a small number but one that is independent of $U$:
\begin{equation}
    U/v_2 \ll |g_{2\sigma}(\ell_f)| = g\ll 1.
\end{equation}
Referring to Fig.~(\ref{fig:flowdiagram}), $g$ corresponds to the radius of the small blue sphere surrounding the non-interacting fixed point. The discussion above implies three distinct behaviors at $\ell_f$, depending on the bare values $U$ and $r$:
\begin{enumerate}
     \item $1 < r \lesssim 7.8 $ and $v_2 \gg U \gg U^{\star}$: All of the couplings $g_a(\ell_f)$ are of order $g$, with the exception of $g_{x\sigma},g_{T}$. As $r \to 1$,
     the ratios of the couplings are increasingly well approximated by their values on the C1S0 ray.
      \item $1 < r \lesssim 7.8 $ and $U \ll U^{\star}$: The couplings other than $g_{2\sigma}$ are suppressed relative to $g$ by $(U/(U^{\star}g))^{\lambda_a}$, with $\lambda_a$ as in Table \ref{tab:lambda}.  (In this sense, the system appears to be approaching the C2S1 ray.)
      \item $7.8 \lesssim r$: 
      The result is formally the same as in case 2 above, but in this case $U^{\star} ~\sim v_2$ 
      and so is not a physically relevant scale for the small $U$ problem.  
\end{enumerate}
From now on, when we say $U \ll U^{\star}$ we mean either case 2 or case 3.  {\em A non-trivial test of the validity of this approach (which we show is satisfied) is that physical quantities we infer at the end of the analysis should not depend on the  value  of $g$ we chose to terminate the first stage RG analysis, so long as $g \ll 1$.}

\section{Analysis of the renormalized Hamiltonian
\label{sec:analysis}}
In this section, we analyze the renormalized Hamiltonian at $\ell = \ell_f$. We start off by switching to a Bosonized representation.

\subsection{Bosonization}
We represent an electron with chirality $p$ ($p = 1$ for $R$, $p=-1$ for $L$), band index $i$, and spin $\alpha$ as
\begin{align}
\psi_{pi\alpha}(x) = 
	\eta_{i\alpha} \sqrt{\frac{\Lambda}{2\pi}} e^{i \sqrt{\pi}\pqty{ \theta_{i\alpha}(x) -p\phi_{i\alpha}(x)}}
	\label{chiral}
\end{align}
where
\begin{equation}
[\phi_{i\alpha}(x),\theta_{i'\alpha'}(x')] = 
	i\Theta (x'-x)\delta_{ii'}\delta_{\alpha\alpha'} \label{eq:commutator},
\end{equation}
$\eta_{i\alpha}$ are Majorana fermions or ``Klein factors'' ensuring that different species of fermion anticommute, and $\Lambda$ is a momentum cutoff. We next make the canonical transformation to spin and charge fields 
\begin{align}
\phi_{i\sigma} &= \frac{1}{\sqrt{2}}\pqty{\phi_{i\uparrow} - \phi_{i\downarrow}} \\
\phi_{i\rho} &= \frac{1}{\sqrt{2}}\pqty{\phi_{i\uparrow} + \phi_{i\downarrow}},
\end{align}
and similarly for $\theta$. 
To complete the bosonized description, we pick the Klein factor ``gauge'' $\eta_{1\uparrow}\eta_{1\downarrow}\eta_{2\downarrow}\eta_{2\uparrow} = 1$. 

The resulting Hamiltonian density  can be expressed as
\begin{equation}
    \mathcal{H} = \mathcal{H}_0 - \sum_i 2\pi v_i g_{i\sigma}\mathcal{H}_{i\sigma}- \pi(v_1+v_2)g_S \mathcal{H}_{S} +\ldots
\end{equation}
where the unperturbed piece is the bosonized version of the non-interacting problem
\begin{equation}
 \mathcal{H}_0 = \sum_{i\nu}
\frac{v_i}{2} \bqty{ (\partial_x \phi_{i\nu})^2 + (\partial_x \theta_{i\nu})^2},
\label{eq:H0}
\end{equation}
the couplings $g_a$ are the renormalized values at $\ell=\ell_f$ 
and
\begin{multline}
\mathcal{H}_{i\sigma} = \\ \frac{1}{8\pi} \bqty{(\partial_x \phi_{i\sigma})^2 - (\partial_x \theta_{i\sigma})^2} 
	-  \pqty{\frac{\Lambda}{2\pi}}^2 \cos(\sqrt{8\pi} \phi_{i\sigma})
	\label{eq:spin_interaction}
\end{multline}
\begin{multline}
    \mathcal{H}_{S} =  \\
    4\pqty{\frac{\Lambda}{2\pi}}^2 \cos\pqty{\sqrt{4\pi}\theta_{-\rho}} 
	\cos\pqty{\sqrt{2\pi}\phi_{1\sigma }} 
	\cos\pqty{\sqrt{2\pi}\phi_{2\sigma }}, \label{eq:pair_tunneling}
\end{multline}
in which $\theta_{\pm \rho} = 1/\sqrt{2}(\theta_{1\rho} \pm \theta_{2\rho})$. 
The terms represented in $\ldots$ above are sinusoidal interactions proportional to $g_{x\sigma}$ and $g_T$, and gradient terms proportional to $g_{x\rho}$ and $g_{i\rho}$. For the present purposes their explicit expressions will not be needed.

\subsection{C1S0 phase for $ U^{\star} \ll U \ll 1$}
Let us briefly review the argument due to BF for a C1S0 phase when $U \gg U^{\star}$. Although $g \ll 1$, 
the  dominant interactions --
including in particular the terms proportional to $g_{1\sigma}$, $g_{2\sigma}$, and $g_{S}$ -- are marginally relevant. As a result, $\phi_{1\sigma}$, $\phi_{2\sigma}$, and $\theta_{-\rho}$ are pinned and the corresponding fluctuational spectrum is gapped. 
Needless to say, the overall charge mode remains gapless, because translation invariance and charge conservation permit only gradient terms for $\phi_{+\rho}$ and $\theta_{+\rho}$. Since the interactions responsible for the gaps all come with a coefficient of order $g$, all gaps are roughly the same size.
Therefore, we refer to this as the single-scale C1S0 regime. Notice also that 
if we now assess the impact of the heretofore neglected subdominant couplings $g_{x\sigma}$, $g_{T}$, we conclude that they are relatively benign in this case, as they are perturbations on top of a maximally gapped state.

\subsection{C1S0 phase for $U \ll U^{\star}$}
To begin with, let us ignore all subdominant interactions, i.e. consider the case in which only $g_{2\sigma}=-g$ is non-zero. Now
$\phi_{2\sigma}$ is governed by the familiar 
$SU(2)$ symmetric sine-Gordon Hamiltonian,
known to yield a gap 
\begin{equation}
    \Delta_{2\sigma}(g) \sim \Lambda e^{-1/g}
\end{equation}
On the other hand, band~1 remains 
non-interacting, resulting in a C2S1 phase.

Now we consider the effect that the subdominant couplings have on this state. While they are indeed parametrically small in $U$, in contrast with the C1S0 phase the C2S1 state contains additional gapless modes which can potentially be gapped out. Here we show that the residual interactions in fact reduce the C2S1 phase to a C1S0 phase.

The intuitive argument is as follows. The singlet pair tunneling interaction, Eq.~(\ref{eq:pair_tunneling}), allows for the 1D analogue of the superconducting proximity effect between the two  bands~\cite{Emery_Kivelson_Zachar_1997,Emery_Kivelson_Zachar_1999}. Thus, band~1 -- which is a metal in the absence of the residual interactions -- inherits a spin gap $\Delta_{1\sigma}$ from band~2. 
 Moreover, the same term acts as an inter-band Josephson coupling, so fluctuations of the relative superconducting phase develops a gap $\Delta_{-\rho}$ as well.  The result is 
a C1S0 phase. 
In this case, however, $\Delta_{1\sigma}$ and $\Delta_{-\rho}$ are subsidiary gaps that vanish in the limit of zero residual interactions, and are therefore suppressed relative to the primary gap $\Delta_{2\sigma}$. We refer to this as the multi-scale C1S0 regime.

We now quantitatively demonstrate this result in a mean-field approximation.
As a first step, we consider the effect of non-zero $g_{S}$ 
 (but continue to ignore the rest of the couplings). Since $\phi_{2\sigma}$ is gapped by an $\order{g}$ interaction, the effect of a parametrically smaller $g_{S} \sim g (U/(U^{\star}g))^{1/4}$ 
will produce correspondingly small changes to its correlations.   Therefore, we can replace the operator $\cos\pqty{\sqrt{2\pi}\phi_{2\sigma }}$ in $\mathcal{H}_{S}$ by its nonzero expectation value, 
 $M(g)$: 
\begin{equation}
    \mathcal{H}_S \to 
    4\pqty{\frac{\Lambda}{2\pi}}^2 M(g) \cos\pqty{\sqrt{4\pi}\theta_{-\rho}} 
	\cos\pqty{\sqrt{2\pi}\phi_{1\sigma }}. 
	\label{eq:mean_field_replacement}
\end{equation}
The above mean-field version of $\mathcal{H}_{S}$ has scaling dimension $3/2$ with respect to $\mathcal{H}_0$, and is therefore a relevant perturbation. 
This indicates an instability of the putative C2S1 phase and results in a pinning of $\phi_{1\sigma}$ and $\theta_{-\rho}$.
Given the scaling dimension $3/2$, the resulting gap magnitudes are
\begin{align}
    \Delta_{1\sigma}(g) \sim \Delta_{-\rho}(g) &\sim  \abs{g_S M(g)}^{2}\Lambda  \\ & \sim  \sqrt{U/U^{\star}} (g^{3/4} M(g))^{2} \Lambda.
    \label{eq:gaps_before_evaluating}
\end{align}
The function $M(g)$ is, like $\Delta_{2\sigma}(g)$, a property of the sine-Gordon theory. We show in Appendix \ref{app:M(g)} that
\begin{equation}
    M(g) \sim g^{-3/4} (\Delta_{2\sigma}(g)/\Lambda)^{1/2} 
\end{equation}
where the leading dependence on $g$, through $\Delta_{2\sigma}(g)$, reflects the scaling dimension of $\cos\pqty{\sqrt{2\pi}\phi_{2\sigma }}$. Inserting this result into (\ref{eq:gaps_before_evaluating}), it follows that the gap ratios are
\begin{eqnarray}
\frac{\Delta_{1\sigma}(g)}{\Delta_{2\sigma}(g)} \sim \frac{\Delta_{-\rho}(g)}{\Delta_{2\sigma}(g)} \sim \sqrt{U/U^{\star}}.
\end{eqnarray}
Note that these are independent of $g$, as required for the consistency of our mean-field approximation.

Now consider the remaining couplings. The interactions proportional to $g_T$ and $g_{x\sigma}$ vanish when we replace functions of $\phi_{2\sigma}$ by their expectation value -- see Appendix \ref{app:remaining_interactions}. The term proportional to $g_{1\sigma}$ as well as the remaining gradient interactions will only lead to quantitative corrections which are parametrically small in $U/(U^{\star}g)$;  again, small perturbations with respect to a maximally gapped phase produce small changes.

The calculations above are for the renormalized Hamiltonian at $\ell = \ell_f$.
Letting $\Delta_{a,0}$ denote the gap for the initial Hamiltonian at $\ell = 0$, we have
\begin{equation}
    \Delta_{a,0} = e^{-\ell_f}\Delta_a(g).
\end{equation}
Using the ray solution to $g_{2\sigma}$ it is straightforward to show that
\begin{equation}
    \ell_f = \ell_{\infty} - 1/g.
\end{equation}
with $\ell_{\infty} = \hat{\ell}_{\infty}/U$ for some $U$-independent $\hat{\ell}_{\infty}$. Consequently, within the one-loop approximation we have that
\begin{align}
    \Delta_{2\sigma,0} &  \sim  e^{- \hat{\ell}_{\infty}/U} \Lambda \label{eq:delta_2sigma}\\
    \Delta_{1\sigma,0} &\sim \Delta_{-\rho,0} \sim  \sqrt{U/U^{\star}} \Delta_{2\sigma,0}
    \label{eq:delta_other}
\end{align}
As required, these expressions are independent of $g$.

The lack of an algebraic pre-factor in Eq.~(\ref{eq:delta_2sigma}) is not be taken seriously, since at two-loop order, $\ell_f$ will acquire corrections logarithmic in $U$. However, Eq.~(\ref{eq:delta_other}) likely remains true despite this modification, as the ratio between different gaps is a property of the renormalized Hamiltonian, not how long it takes to flow there. 

\subsection{Power law CDW and SC correlations \label{sec:luttinger_K}}
Regardless of which regime is being considered, the charge density wave (CDW) and singlet superconducting (SC) correlations decay algebraically reflecting the existence of a gapless overall charge mode. Writing the low energy effective action for $\phi_{+\rho}$ as 
\begin{equation}
S_{\text{eff}} = \int\dd x \dd \tau
\frac{1}{2K} \bqty{(v \partial_x \phi_{+\rho})^2 + \frac{1}{v} (\partial_{\tau}\phi_{+\rho})^2},
\label{2K}
\end{equation}
we deduce that there exist  CDW and  SC correlations that decay as $|x|^{- 2K}$ and $|x|^{-1/(2K)}$, respectively. The SC correlations can be identified with the usual BCS-type pairing between time-reversed pairs;  this power law characterizes the long-distance correlations of $O_{iS}$ (Eq.~\ref{SCOP}) for either band, $i=1$ or $2$. However, the charge density correlations -- which oscillate with a CDW ordering vector $Q=2(k_{F1}+k_{F2})$ -- are not related in any way to the Peierls-like CDW order associated with the Fermi surface nesting vectors, $2k_{Fi}$, or $k_{F1} \pm k_{F2}$.  Correspondingly, expressed in terms of chiral fermionic fields (Eq.~\ref{chiral}), the quasi-long-range CDW correlations are properties of  the composite density operator, 
\begin{equation}
    J_{\text{comp}} = \sum_{\alpha,\alpha^\prime} \left[\psi_{L,1,\alpha}^\dagger \psi_{L,2,\alpha^\prime}^\dagger\psi_{R,2,\alpha^\prime}\psi_{R,1,\alpha}+ \hc\right].
\end{equation}
It should be stressed that despite the weak coupling approach we have taken, this is an intrinsically strong coupling result that cannot be inferred directly from the non-interacting electronic structure.

The relative strength of the CDW and SC correlations is determined by the precise value of $K$.  At least two distinct aspects of the solution affect this.  Firstly, $\phi_{+\rho}$ mixes modes with different velocities. Secondly, $g_{i \rho}$ and $g_{x\rho}$ explicitly renormalize the gradient terms.
The case $U \gg U^{\star}$ was considered by BF. In this regime, both effects mentioned above are important, as $g_{i \rho} \sim g_{x\rho} \sim g$. Consequently, any simple approximation for $K$ will explicitly depend on  $g$, i.e. on the point at which the RG flows are terminated. For this reason, BF cautioned against taking too seriously their estimate for $K$, and noted only that it tends to increase with $|r-1|$. 
(They give an account of this in Appendix~B of Ref.~\onlinecite{Balents_Fisher_1996}.) 

Here, we consider $U \ll U^{\star}$, where to leading order only the mixing of modes with different velocities is important. In the harmonic approximation, we find
\begin{equation}
    K = \sqrt{2+r+r^{-1}}/2 + \ldots
\end{equation}
where $\ldots$ denotes terms parametrically small in $U/(gU^{\star})$. Note that to leading order, this result is independent of $g$. 
Notice also that $K > 1$, 
which implies that the SC susceptibility diverges  as $T\to 0$ whereas the CDW susceptibility remains finite.  This state is as close to a superconductor as a 1D system can be.

\section{Global RG flow \label{sec:global_RG}}
Our strategy so far has been to solve the renormalized theory -- defined at a point in the RG flow where we still have perturbative control -- using some reasonable approximations. In particular, we have not attempted to follow the flows out to the strong coupling
C1S0 and C2S1  fixed lines. However, it is worth asking what sort of global RG flow is consistent with our results.

The simplest possibility is illustrated in Fig.~(\ref{fig:flowdiagram}), which contains for some fixed $r \lesssim 7.8$ a projection onto the $(g_{2\sigma},g_{S},K)$ subspace
of those flow lines 
corresponding to various small $U$ Hubbard initial conditions. The C1S0 
fixed line represents a critical phase with  continuously varying critical exponents  parameterized by the stiffness $K$,  described above, 
while the C2S1 fixed line is really a ``fixed hyper-surface'' in interaction space parameterized by multiple gradient parameters. The ball of radius $g$ about the origin (i.e. the free fermion fixed point) contains the 
complex flows discussed in Sec.~\ref{sec:RG_analysis}.

The flow line for each initial condition, corresponding to a given value of $U$, emerges from this ball pointing in different directions. 
For $U \gg U^{\star}$, the 
initial 
flow is already in the direction of the C1S0 fixed line. 
As illustrated, it is reasonable to assume that this continues all the way to this fixed line.
For $U \ll U^{\star}$, on the other hand, the 
initial flows are toward the C2S1 fixed 
line.  
However, 
the above analysis implies that in the relevant range, this fixed line is itself perturbatively unstable, so the flows ultimately bend away and eventually also approach the C1S0 line.

Finally, note that in accordance with our discussion in Sec.~\ref{sec:luttinger_K}, depending on $U$ the flow likely actually terminates at different points (i.e. different values of $K$) along the C1S0 fixed line.

\section{DMRG study of the two-leg ladder at $U=4$ \label{sec:DMRG}}

We now report the results of a density matrix renormalization group (DMRG) study of the two-leg ladder at $U = 4$.  In agreement with both the weak coupling theory and with previous DMRG studies  \cite{Noack_White_Scalapino_1996,White1997,White1998,White1999,White2009,Scalapino2012,Dolfi_Bauer_Keller_Troyer_2015,Dodaro2017,Jiang2018tj, Jiang2019} at $U \geq 8$, we find that the ground state is a C1S0 Luther-Emery liquid.
Specifically, as we will show below, we find power-law SC and CDW correlations, exponentially falling spin correlations, and a central charge $c=1$. While the case $U = 4$ is by no means a ``weak'' interaction, we will see that certain aspects of the solution are best understood from a weak-coupling perspective.

Before we present these results, it is worth explicitly noting why DMRG calculations at small $U$ are so challenging: the number of block states needed to faithfully represent the ground state grows rapidly with the correlation length, and as $U \to 0$, the correlation length diverges exponentially with $1/U$. Even for the relatively simple case of the two-leg ladder, and keeping $24,000$ effective $U(1)$ block-states, $U=4$ is the smallest interaction strength for which we have been able to obtain reliable results.

\subsection{Results}

Unless explicitly stated otherwise we work with the following parameters, in addition to $U=4$. First, we set the inter-chain hopping $t_{\perp} =1$; i.e. we set it equal to the rung hopping, which is already set to $1$. Letting $\delta = 1-n$ denote hole doping, where $n$ is the electron density per site, we work at $\delta = 1/12$. The system size is $L_x = 192$. We keep up to $24,000$ effective $U(1)$ block-states and extrapolate all quantities to the zero truncation error limit. Additional calculational details can be found in Appendix~\ref{app:numerical_details}.

Below, $c_{xj\alpha}$ will refer to the operator which annihilates a spin-$\alpha$ electron at position $x$ along leg $j$ (this is the same convention used in Sec.~(\ref{sec:Hamiltonian})). The position $x$ begins at $x = 1$ on the left edge. Also, $\bm{S}_{xj} = (1/2)c_{xj\alpha}\bm{\sigma}_{\alpha\beta}c_{xj\beta}$ will denote the spin operator at  position $x$ along leg $j$.

\begin{figure}
    \centering
    \includegraphics[width=8.5 cm]{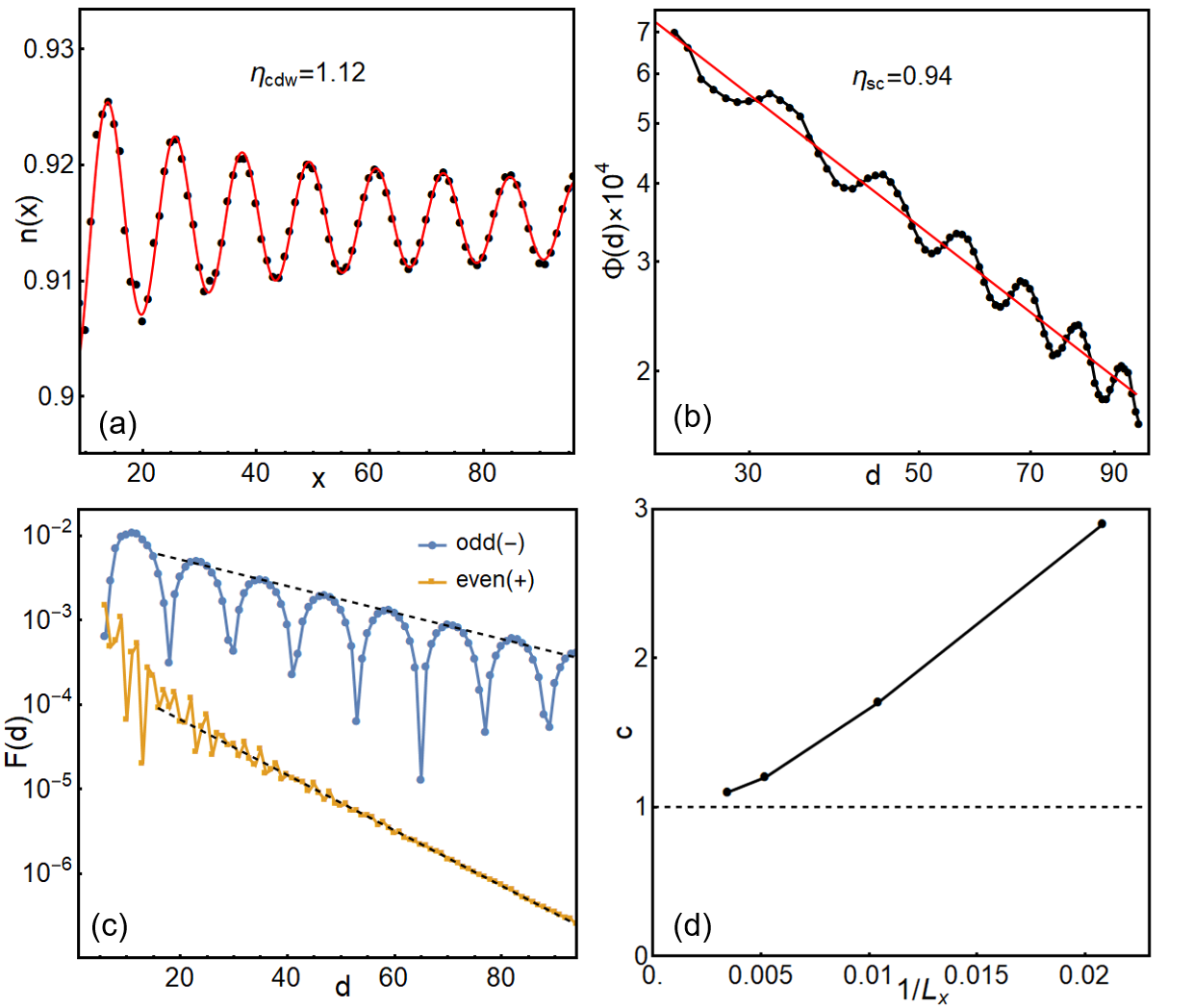}
    \caption{
    Physical properties of the two-leg ladder for $U=4$, $t_{\perp}=1$, and $\delta = 1/12$. In (a)-(c), $L_x = 192$. (a) Density profile $n(x)$ for the left half of the ladder. The red line is a fit to a power-law decaying oscillation with exponent $\eta_{\text{CDW}}/2$, where $\eta_{\text{CDW}} = 1.12$; (b) Power-law decaying pair-pair correlation $\Phi(d)$, with exponent $\eta_{\text{SC}}=0.94$; (c) Exponentially decaying spin-spin correlations $F_{\pm}(d)$; (d) The central charge extracted from systems with length $L_x=48, 96, 192$ and $288$.
    }
    \label{fig:DMRG}
\end{figure}

In Fig.~\ref{fig:DMRG}, we exhibit
several  properties of the C1S0 phase. We begin with the CDW and SC correlations, expected to fall off as power laws in the C1S0 phase. In Fig~\ref{fig:DMRG}(a) we show $n(x)$, the expectation value of the electron density at site $x$. Due to the open boundary at $x=0$, $n(x)$ contains an oscillating component. In a C1S0 phase, we expect these oscillations to decay as a power law, with an exponent equal to half the exponent governing the fall-off in CDW correlations \cite{White_Affleck_Scalapino_2002}. We indeed find a power law form $x^{-\eta_{\text{CDW}}/2}$ for the amplitude of the oscillations, with $\eta_{\text{CDW}} = 1.12(4)$. Next, in Fig.~\ref{fig:DMRG}(b) we show the SC correlation function 
\begin{equation}
    \Phi(d)=\langle\Delta_Y^\dagger(x_0)\Delta_Y(x_0+d)\rangle
\end{equation}
where
\begin{eqnarray}
\Delta_Y(x)=\frac{1}{\sqrt{2}}(c_{x,1,\uparrow}^\dagger c_{x,2,\downarrow}^\dagger-c_{x,1,\downarrow}^\dagger c_{x,2,\uparrow}^\dagger)
\end{eqnarray}
creates a vertically ($Y$ direction) oriented Cooper pair on rung $x$, and the reference rung $x_0$ is set to $L_x/4$. We find $\Phi(d) \sim d^{-\eta_{\text{SC}}}$ with $\eta_{\text{SC}} =0.94(2)$. The product 
\begin{eqnarray}
\eta_{\text{CDW}}\cdot \eta_{\text{SC}}=1.05(6)
\end{eqnarray}
is within error bar of the theoretically expected value $1$. The corresponding Luttinger parameter, defined in Sec.~(\ref{sec:luttinger_K}) is $K \approx 0.5$.  Finally, to determine the nature of the the pairing, we have computed the SC correlation function between a vertically oriented Cooper pair and a horizontally oriented Cooper pair. We find that it is negative, indicating $d$-wave-like pairing. 

Next we examine the spin correlations. In Fig.~\ref{fig:DMRG}(c), we plot the correlation functions $F_{\pm}(d)$, defined as:
\begin{equation}
F_{\pm}(d)=\frac{1}{4}\ev{(\bm{S}_{x_0,1} \pm \bm{S}_{x_0,2}) \vdot (\bm{S}_{x_0+d,1} \pm \bm{S}_{x_0+d,2})}.
\end{equation}
That is, $\pm$ refers to the correlation function for the combination of spin operators which is even/odd under exchanging the two legs. As in the SC correlation function, we set $x_0 = L_x/4$. We find an exponential decay, $F_\pm(d) \sim e^{-\xi_{\pm}d}$, with $\xi_+=13.3$ and $\xi_-=27.5$. The fact that the correlation lengths are significantly longer than the lattice spacing means the system is not too far from the free fermion critical point. In contrast with the case $U=4$, for $U=8$ we find shorter correlation lengths, $\xi_+=4.6$ and $\xi_-=9.6$. It should be noted that the appearance of two distinct correlation lengths is 
 apparently
 unrelated
to the hierarchy of gap scales discussed above, as the ratio $\xi_{-}/\xi_{+} \approx 2$ is essentially unaffected by changing $U$.

In Fig.~\ref{fig:DMRG}(d) we show the measured central charge for several different system sizes $L_x$.  For the two longest systems, it approaches $c = 1$. Taken together with the behavior of the spin and charge correlations, this implies that the only gapless mode is the overall charge mode. Notice that $L_x$ must be appreciably larger than the correlation length to accurately determine the central charge; at the smallest system size $L_x = 48$ (already twice the longest correlation length) its apparent value is larger, $c \approx 3$.

\begin{figure}
    \centering
    \includegraphics[width=0.8\linewidth]{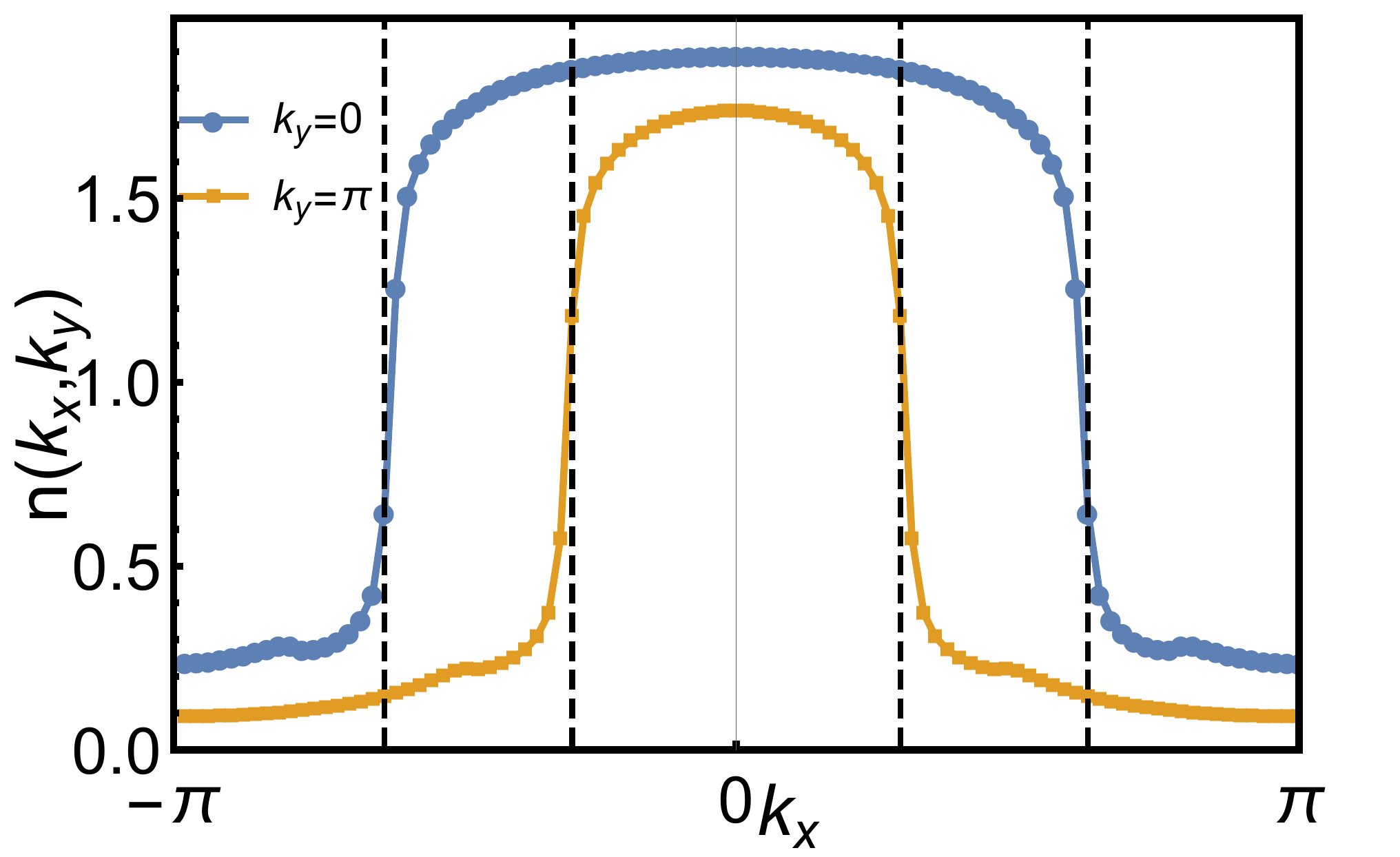}
    \caption{Electron density $n(\bm{k})$ in momentum space. Dashed lines indicate the Fermi points of non-interacting bands.
    \label{fig:n(k)}}
\end{figure}
Finally, in Fig \ref{fig:n(k)}, we plot the single particle occupancy in momentum space, $n(\bm{k})$, defined by
\begin{multline}
    n(\bm{k}) = \\
    \frac{1}{2L_x} \sum_{x,x',j,j',\alpha} \ev{c^{\dagger}_{x,j,\alpha}c_{x',j',\alpha}}e^{i k_x(x-x') +ik_y(j-j')}.
\end{multline}
Here $k_y = 0$ ($\pi$) is the bonding (anti-bonding) band.
Although all single particle excitations are gapped in the C1S0 phase, $n(\bm{k})$ nevertheless exhibits an abrupt drop near the Fermi points of the non-interacting system. 
Note that if we did not have access to such accurate data on such long systems, it would be tempting to interpret the behavior of $n(\bm{k})$ as evidence that the non-interacting C2S2 phase survives for a finite range of $U$.
In contrast with $U=4$, when we increase the interaction strength to $U = 12$, we find that the features of $n(\bm{k})$ are highly rounded and not even centered about the Fermi points of the non-interacting system (see Appendix \ref{app:numerical_details}).

\section{The Triangular Lattice \label{sec:triangular}}

To illustrate the usefulness of the present approach, we 
sketch its application  to the case of the triangular lattice Hubbard model at half-filling.  (A more complete study will be reported in a future publication \cite{Gannot_Jiang_Kivelson_2020}.)  This problem has recently been studied at intermediate to large $U$ by DMRG methods \cite{Szasz_et_al_2020,Shirakawa_et_al_2017}. 
For  $U> U_{c1} \approx 10 t$, these studies show an insulating phase that is a 1D version of the three-sublattice $120^{\circ}$ magnetically ordered state believed to be the ground-state of the 2D spin 1/2 Heisenberg antiferromagnet.  Intriguingly, for $U_{c2} < U < U_{c1}$ with $U_{c2} \approx 8 t$, a distinct intermediate insulating phase is observed, 
which has been conjectured to reflect the existence of 
a spin-liquid phase in the 2D limit. 
Depending on cylinder geometry and/or computational details, these studies have adduced evidence that the spin-liquid in question is either 
fully gapped and chiral \cite{Szasz_et_al_2020}, or gapless and non-chiral \cite{Shirakawa_et_al_2017}.
Finally, for $U< U_{c2}$, a conducting phase appears which has been identified as ``metallic,'' i.e. to have the same number of gapless modes as in the 
$U=0$ limit.

The 
small $U$ approach explored in the present paper is clearly of limited  use for giving  insight into the nature of the phases that occur for $U>U_{c2}$, but if it is true that a single phase arises in the range $0<U<U_{c2}$, then insight into the nature of this phase can be obtained by analyzing the small $U$ limit.  In 2D, it has already been shown \cite{Raghu_Kivelson_Scalapino_2010,Nandkishore_Thomale_Chubukov_2014} that the ground-state of the triangular Hubbard model at small $U$ and $n=1$ is a $d+id$ superconducting state.  Presumably, for a cylinder of large circumference, this would correspond to a C1S0 phase with a broken discrete symmetry, i.e. it would be a chiral Luther-Emery liquid.  But for the small  circumference cylinders actually studied by DMRG, the correct comparison should be based directly on a multicomponent 1DEG as in the present paper.  We thus conclude by applying the insights obtained from the present study 
to the four-leg triangular lattice cylinder, referred to in the DMRG literature as YC4 and studied, among other cylinders, in Ref.~\onlinecite{Szasz_et_al_2020}.

\begin{figure}
\centering
\includegraphics[width=0.50\linewidth]{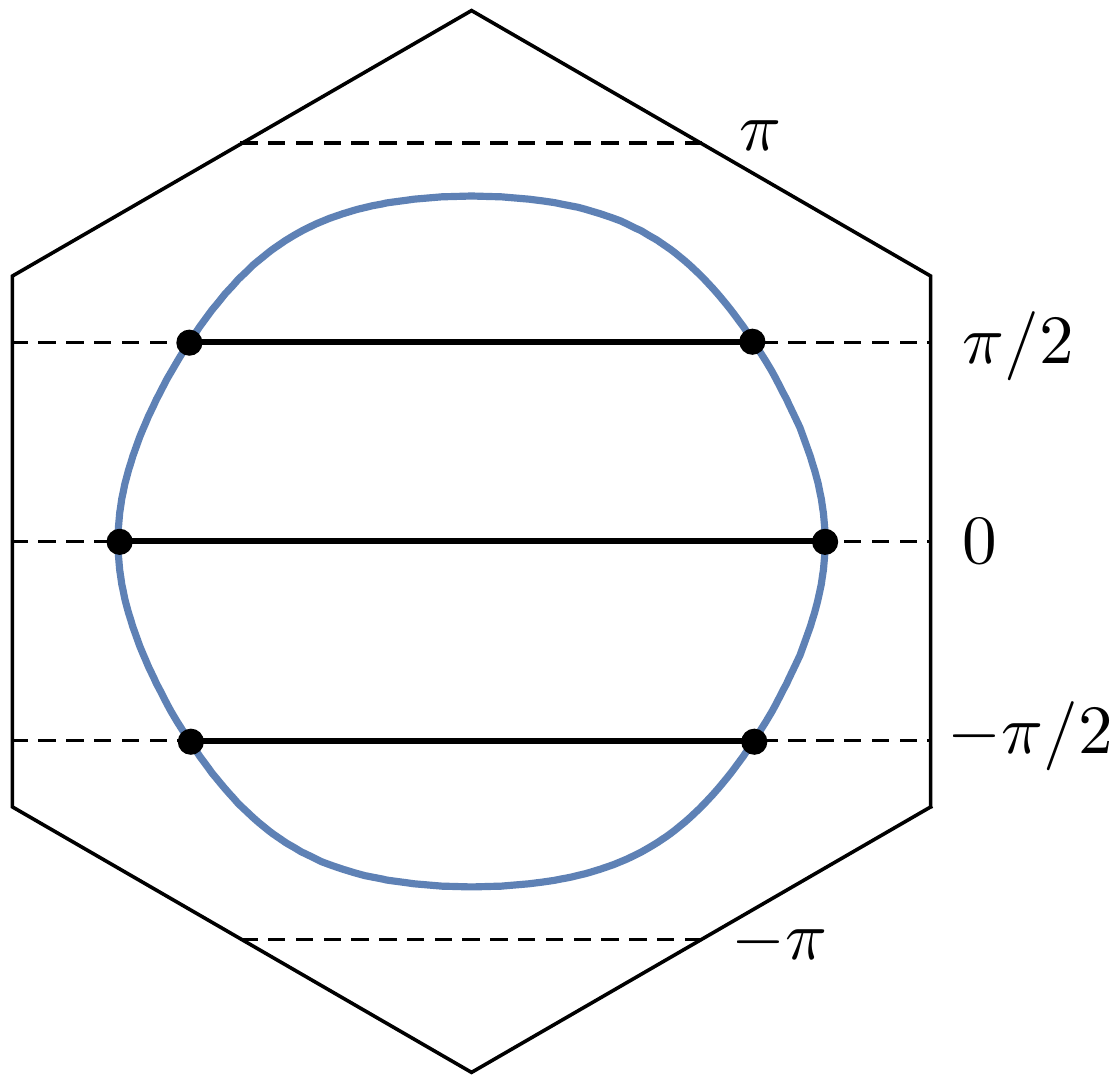}
\caption{Fermi surface structure of the $U=0$ triangular lattice Hubbard model on a YC4 cylinder. The allowed momenta are the slices $k_y = 0, \pm \pi/2, \pm \pi$ through the 2D Brillouin zone. The thick black lines indicate states occupied at $n=1$, and the black dots indicate the Fermi points. The blue curve is the Fermi surface of the 2D system at the same Fermi energy. 
\label{fig:triangularband}}
\end{figure}

The YC4 cylinder band structure is obtained by restricting the two-dimensional triangular lattice band structure to transverse momenta $k_y  
=0,\ \pm \pi/2, \ $ and $\pi$.  
When $n=1$, the bands with $k_y = 0$ and $ \pm \pi/2$ cross the Fermi energy, as shown in Fig.~\ref{fig:triangularband}. The Fermi velocities, $v_i$, for the bands with $k_y=\pm \pi/2$ (indexed as $i=\pm 1$) are equal due to reflection symmetry in $y$. Interestingly, however,
for $n=1$ (half-filling) all of the Fermi  velocities are equal: $v_0 = v_{\pm 1}$ where $i=0$ refers to the band with $k_y=0$; this is not the consequence of any symmetry, and hence would not be true in slightly modified versions of the model, e.g. if a small amount of second-neighbor hopping were included.

To study the properties of this cylinder for small but non-zero $U$, we can employ the multi-band version of the perturbative RG equations, derived and studied in detail by Lin, Balents, and Fisher \cite{Lin_Balents_Fisher_1997}.  Defining $
r \equiv v_0/v_{\pm 1}$, when $
r= 1$ these equations are fully symmetric under exchanging any two bands. In the naive continuum limit, so too are the initial conditions. Again,  this is not an exact symmetry of the lattice Hamiltonian, but deviations from this symmetry vanish as $U \to 0^{+}$.

The couplings at this fully symmetric point flow out along a ray which is analogous to the C1S0 ray found in the two-leg ladder at $r=1$.  
Specifically, upon bosonizing the problem, the  interactions that grow along  the outgoing ray that can lead to the  opening of gaps are the intra-band spin interactions
 $g_{\pm 1 \sigma}$ and  $g_{0\sigma}$ (analogues of $g_{1\sigma}$ and $g_{2\sigma}$ in the two-leg ladder) and the pair tunnelling terms,
 $g_{\tilde{\imath}S}$, (analogous to $g_S$ in the two-leg ladder), where $\tilde{\imath}=\pm 1$ refers to  pair-tunnelling between the $k_y=0$ and $k_y=\pm \pi/2$ bands, while  $\tilde{\imath}=0$ refers to pair-tunnelling between the $k_y=-\pi/2$ and $k_y=+\pi/2$ bands.\footnote{
 We are specifying here the $k_y$ of the right mover in each Cooper pair; the left mover is understood to have opposite $k_y$.}
 
 Significantly, the peculiar symmetry of this ray ensures that all the bands are interchangeable, i.e. that $g_{i\sigma}=g_\sigma$ and $g_{\tilde{\imath}S}=g_S$ for all $i$ and $\tilde{\imath}$.  A growing $g_{\sigma}$ is easily seen to   gap out all the spin modes. Letting $\Theta_i$ denote the phase of superconducting order parameter on band $i$, the latter acts as a Josephson coupling
\begin{equation}
\mathcal{H}_J  = -J 
\sum_{i < j}\cos(\Theta_i -\Theta_{j})
\end{equation}
with $J \propto g_S<0$. 
(The proportionality constant depends on the expectation value of appropriate functions of the spin fields.)
Up to an overall shift of the total phase, the pattern of $\Theta_\alpha$'s that minimizes this expression either increases by $2\pi/3$ each time one moves clockwise between two Fermi points, or else decreases by $2\pi/3$. 
This winding breaks time reversal and $x$ and $y$ mirror symmetries.  The result is 
a chiral C1S0 (Luther-Emery) phase. 

The chiral C1S0 (C1S0-$\chi$) phase is an especially attractive candidate for the small-$U$ phase on the YC4 cylinder because there is a natural mechanism for transitioning from it into 
a fully-gapped chiral spin-liquid.
At half filling, there are a variety of six-fermion umklapp scattering terms that are allowed, such as
\begin{eqnarray}
H_{\text{umk}} = \int dx \ g_{\text{umk}} \prod_{i}\left[\sum_\alpha\psi_{R,i,\alpha}^{\dagger}\psi_{L,i,\alpha}\right] + \hc
\label{eq:umklapp}
\end{eqnarray}
The bare value of such terms is zero in the original Hubbard model, and they are manifestly irrelevant at the $U=0$ fixed point.  However, 
 on the strong coupling C1S0 fixed line, its dimension depends on the  Luttinger exponent $K$. If $K$ changes with increasing $U$ in such a way that the umklapp term becomes relevant (namely, $K$ drops below $4/3$) then the overall charge mode becomes gapped, yielding a chiral 
 insulator. The transition would be in the Kosterlitz-Thouless universality class. Such a state can naturally be identified as the finite cylinder descendant of a fully gapped chiral spin-liquid.

The C1S0-$\chi$ ray is the true asymptotic ray only at exactly  $r=1$, reflecting the fact that the symmetry between the three bands is not generic. However, at finite $U$ the flow starts to deviate from this ray only after the couplings get so large that the perturbative approach 
breaks down. A preliminary analysis suggests the C1S0-$\chi$ phase occupies a fan in the $(r,U)$ plane, emerging from $r=1$. For $U \ll 1$, the width of this fan never exceeds $10^{-4}$ in $r$.

Given the narrowness of this fan, at finite $U$ more general values of the parameter $r$ may be relevant. Consider first $r \lesssim 1$. The asymptotic rays are as follows:
\begin{itemize}
    \item For $0.54 < r < 1$: C3S2, along which
    band 0 is span gapped but bands $\pm 1$ are not.
    \item For $r< 0.54$: $s$-wave-like C2S1, along which bands $\pm 1$ are spin gapped with the same SC phase, but band $0$ is not.
    \item At the isolated point $r=0.54$: non-chiral, $d$-wave-like C1S0 ray (C1S0-$d$), along which all bands are spin gapped, and the SC phase on band $0$ is opposite the SC phase on bands $\pm 1$.
\end{itemize}
However, the C1S0-$d$ ray is only very weakly unstable away from its isolated point of stability, and is in this sense analogous to the C1S0 ray in the two-leg ladder. By the same reasoning used in the two-leg ladder, it follows that unless $U$ is smaller than some extremely small crossover scale $U^{\star}$, it is the weakly unstable C1S0-$d$ ray which directly determines the phase of matter for all $r \lesssim 1$. Also in analogy with the two-leg ladder, for $U \lesssim U^{\star}$ pair tunneling destabilizes putative C3S2 and $s$-wave-like C2S1 phases, resulting again in a C1S0-$d$ phase.

The upshot of this analysis is a ubiquitous C1S0-$d$ phase for $r \lesssim 1$. This implies that the left-hand boundary of the fan containing the C1S0-$\chi$ phase is a true phase boundary; this is in contrast with the crossover scale $U^{\star}$ which here and in the two-leg ladder separates single-scale from multi-scale regimes. If the charge mode of the C1S0-$d$ phase is gapped by the umklapp interaction (\ref{eq:umklapp}), the resulting insulator can naturally be identified as the finite cylinder descendant of a non-chiral, $Z_2$ spin liquid.

Finally, for $r > 1$ the asymptotic ray describes a $d$-wave-like C2S1 phase in which bands $\pm 1$ are spin-gapped with opposite SC phase. In contrast with $r < 1$, the range $r > 1$ contains no weakly unstable C1S0 ray.  Moreover, even when we consider the effect of pair tunneling to the remaining un-gapped band, the spin gap proximity effect with the two gapped bands interfere destructively.  The ultimate fixed point in this case is an interesting question which we reserve for future study.

An unambiguous result, however, is that at least one mode is gapped for $U \ll 1$. That is, the phase adjacent to $U = 0$ has central charge $c < 6$. This is in disagreement with Ref.~\onlinecite{Szasz_et_al_2020} which reports $c=6$ for $U<U_{c2}$.
Ref.~\onlinecite{Szasz_et_al_2020} also reports an 
apparent singularity in the single particle occupancy $n(\bm{k})$ as evidence for a maximally gapless metallic state. There are two possible explanations behind this disagreement. First, there may be an additional phase transition at some $U_{c3}$ which is small but inaccessible by weak coupling. We believe a more likely explanation is that there is indeed a single phase with $U<U_{c2}$, but that it is not the C3S3 phase suggested by the DMRG studies. 
Because all of the
gaps vanish exponentially as $U \to 0$, it is intrinsically difficult to distinguish  gapped modes from gapless ones 
in any numerical study.
Similarly, when the correlation length is long $n(\bm{k})$ can be a misleading diagnostic for gaplessness.

Our DMRG results for the two-leg ladder at $U=4$ illustrate these challenges. In the thermodynamic limit this system has a central charge of $c=1$, and is as gapped as possible. However, as shown in Fig.~\ref{fig:DMRG}(d) the measured central charge is apparently $3$ at smaller system sizes, decreasing to the true asymptotic value $c=1$ only for much larger systems. 
This illustrates the difficulties in extracting the true central charge from numerics.
Moreover, the single particle occupancy $n(\bm{k})$ in Fig.~\ref{fig:n(k)} exhibits a rapid 
drop that superficially appears non-analytic at the non-interacting $k_F$; had we not already established that the single-particle Green function falls exponentially with distance, it would have been tempting to follow the reasoning of Ref.~\onlinecite{Szasz_et_al_2020} and to interpret this as evidence of a metallic state.

\begin{acknowledgments}
We thank Xiao-Qi Sun and Yoni Schattner for helpful discussions.  We would like to thank Sophia Kivelson for  Fig.~\ref{fig:flowdiagram}. This work was supported in part by Department of Energy, Office of Basic Energy Sciences, under Contract No. DEAC02-76SF00515 at Stanford.
Parts of the computing for this project were performed on the Sherlock cluster.  
\end{acknowledgments}

\appendix
\section{Conventions and RG equations \label{app:RG}}
Our convention for the couplings differs slightly from BF. First, in place of their $g_{t\rho}$ and $g_{t\sigma}$,
we use $g_{S} = g_{t\rho} - \frac{3}{4}g_{t\sigma}$ and $g_{T} = g_{t\rho} + \frac{1}{4}g_{t\sigma}$ (with the same relation holding for the tilded versions of these couplings). Second, our $g_{i\rho}$,  $g_{i\sigma}$ contain an additional factor of $(v_1+v_2)/(2v_i)$. Introducing 
\begin{equation}
\gamma = (v_1+v_2)^2/(4v_1 v_2) = (2+r+r^{-1})/4 \label{eq:gamma}
\end{equation}
the one-loop RG equations in our notation are:
\begin{subequations}
\label{eq:RG}
\begin{align}
&\dv{{g}_{1\rho}}{\ell} = \dv{{g}_{2\rho}}{\ell} = - \gamma \dv{{g}_{x\rho}}{\ell} = \gamma \pqty{\frac{1}{4}g_S^2 + \frac{3}{4} g_T^2} \\
&\dv{{g}_{1\sigma}}{\ell} = -g_{1\sigma}^2 -\gamma(g_{S}^2 -g_{T}^2) \\
&\dv{{g}_{2\sigma}}{\ell} = -g_{2\sigma}^2 -\gamma(g_{S}^2 -g_{T}^2) \\
&\dv{{g}_{x\sigma}}{\ell} = -g_{x\sigma}^2 +2 g_{T}(g_S+g_T) \\
&\dv{{g}_{S}}{\ell} = g_{S}(g_{1\rho} + g_{2\rho} -2 g_{x\rho}) - \frac{3}{4}g_S(g_{1\sigma}+g_{2\sigma}) \nonumber \\
&\qquad + \frac{3}{2} g_{T}g_{x\sigma} \label{eq:g_S_beta} \\
&\dv{{g}_{T}}{\ell} = g_{S}(g_{1\rho} + g_{2\rho} -2 g_{x\rho}) + \frac{1}{4}g_T(g_{1\sigma}+g_{2\sigma} -4 g_{x\sigma})  \nonumber \\ &\qquad + \frac{1}{2} g_{S}g_{x\sigma}.
\end{align}
\end{subequations}

\section{Additional analysis of the RG flows}
\label{app:RG_details}
\subsection{C1S0 ray}
Here we give explicit expressions for the $G_a$ of the C1S0 ray. With $\gamma$ be as in (\ref{eq:gamma}),
\begin{subequations}
\begin{align}
    &G_{1\rho} = G_{2\rho} = -\gamma G_{x\rho} = \frac{\gamma}{-1+8\gamma+3\sqrt{1+8\gamma^2}} \\
    &G_{1\sigma}=G_{2\sigma} = -\frac{1+4\gamma-\sqrt{1+8\gamma^2}}{2(1+\gamma)} \\
    &G_{x\sigma} = 0 \\
    &G_{S} = -\frac{2}{\sqrt{-1+8\gamma+3\sqrt{1+8\gamma^2}}} \\
    &G_{T} = 0
\end{align}
\end{subequations}

\subsection{Plots of the remaining coupling ratios}
\begin{figure}
    \centering
    \includegraphics[width=\linewidth ]{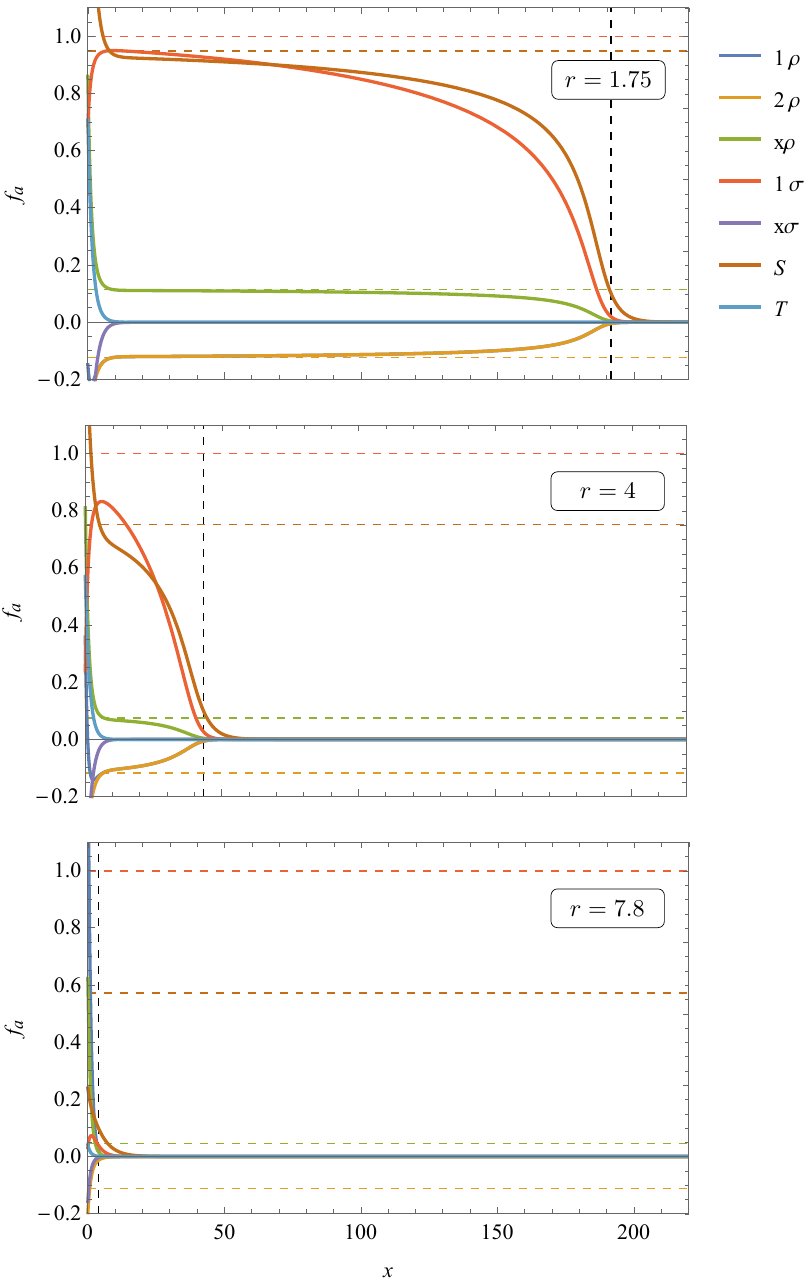}
    \caption{
    Ratios $f_a(x)$. Each panel corresponds to a fixed value of $r$. Within each panel, the different colors represent  different $f_a$. For each $f_a$, the horizontal dashed line of the appropriate color gives the value of $f_a$ on the C1S0 ray. The black vertical dashed line gives $x^{\star}$ as defined in Eq.~(\ref{eq:x_star_def}).
    \label{fig:appendix_ratios}}
\end{figure}

In Fig.~\ref{fig:appendix_ratios}, we show the ratios $f_a(x)$ for several values of $r$. Each panel corresponds to a fixed value of $r$, and different colors represent different ratios $f_a$.

\subsection{Divergence of $x^{\star}$}
\begin{figure}
    \centering
    \includegraphics[width=0.9\linewidth ]{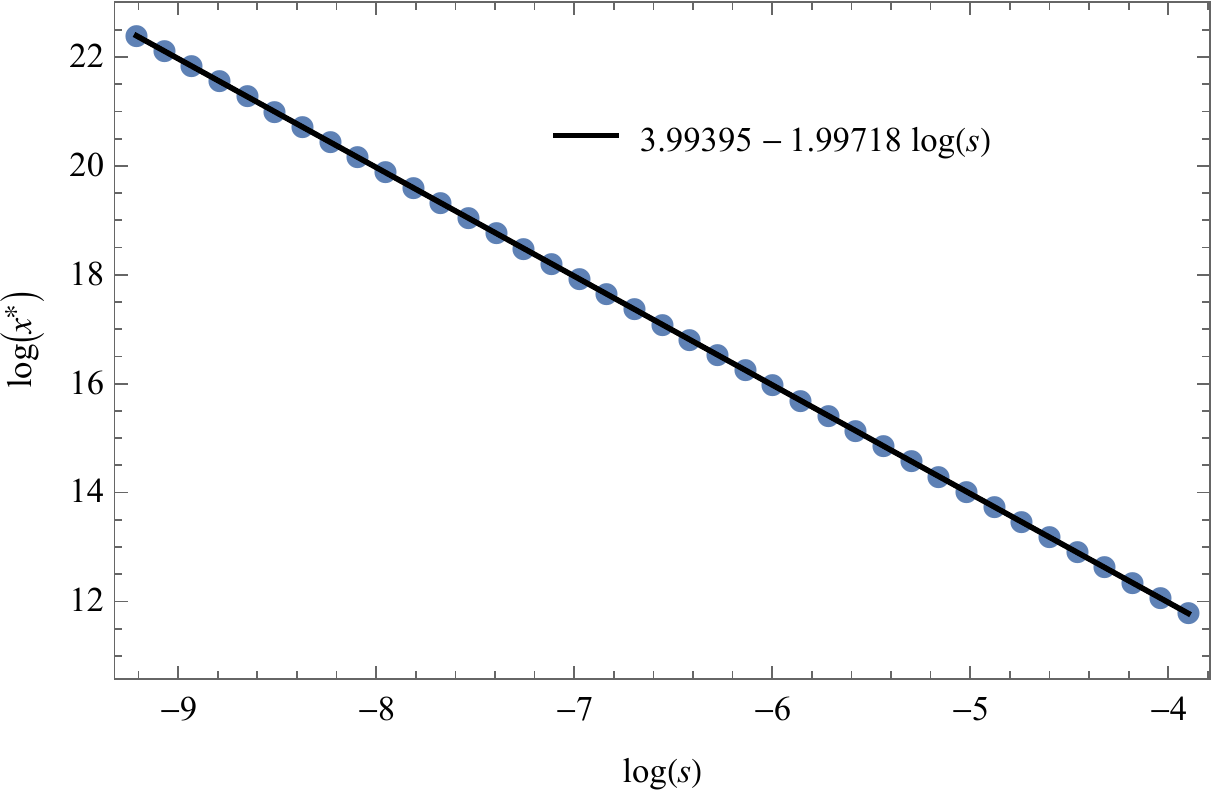}
    \caption{Numerically determined $\log(x^{\star})$ versus $\log(s)$ (the blue circles) together with a linear fit.
    \label{fig:divergence}}
\end{figure}

Here, we show that $x^{\star}$ diverges with $s \equiv r-1$ as $x^{\star} \sim s^{-2}$. It turns out to be somewhat subtle to analyze this divergence analytically, so we do so numerically. In Fig.~(\ref{fig:divergence}) we plot the numerically determined values of $\log(x^{\star})$ versus $\log(s)$, and find a near-perfect fit to $\log(x^{\star}) = a_0 + a_1\log(s)$ with $a_1 \approx -2.00$, consistent with $x^{\star} \sim s^{-2}$.

\subsection{Exponents $\lambda_a$}
\begin{figure}
    \centering
    \includegraphics[width=\linewidth ]{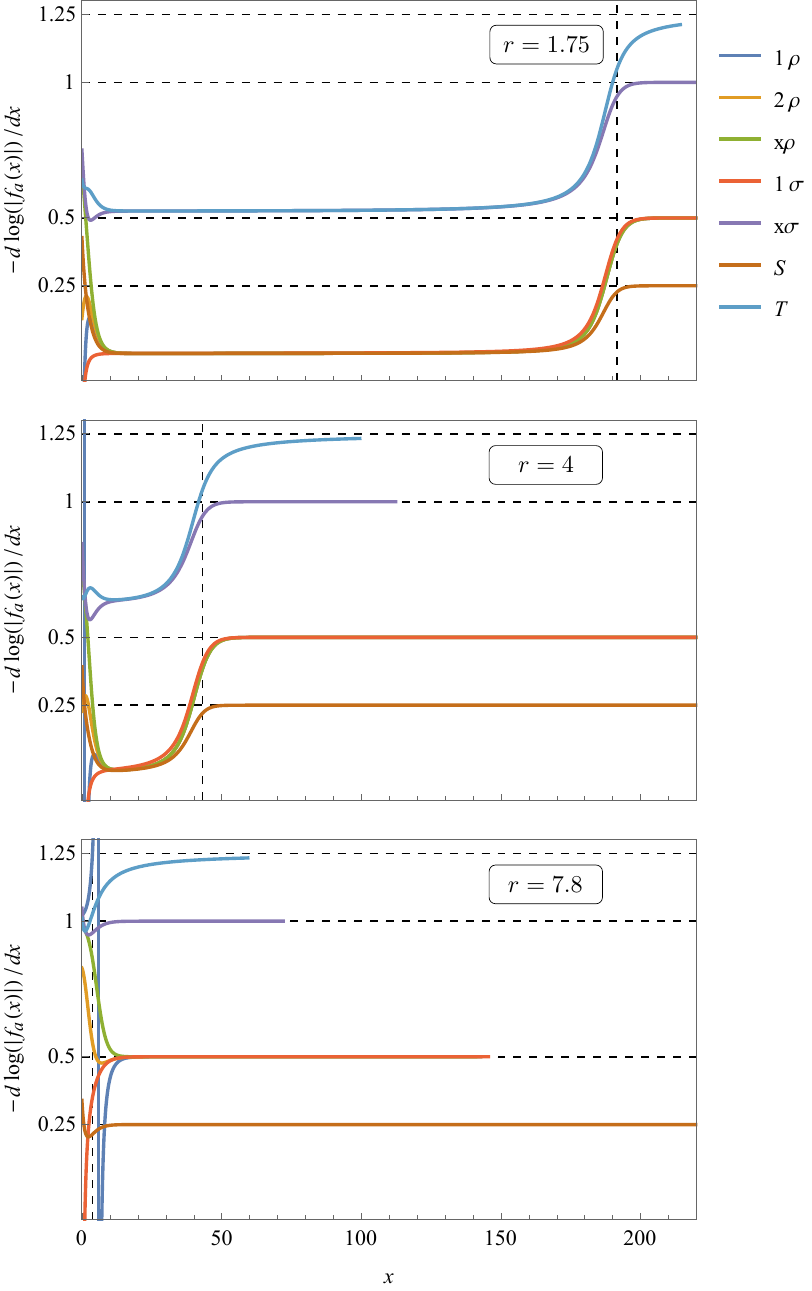}
    \caption{
    Plots of the functions $ -\dv*{\log(|f_a(x)|)}{x}$. Each panel corresponds to a fixed value of $r$. Within each panel, the different colors represent the different indices $a$. The black vertical dashed line gives $x^{\star}$ as defined in Eq.~(\ref{eq:x_star_def}). Some curves are cutoff once the corresponding $f_a$ is numerically too small to give reliable results.
    \label{fig:appendix_exponents}}
\end{figure}

In Fig.~(\ref{fig:appendix_exponents}), we plot the functions $ -\dv*{\log(|f_a(x)|)}{x}$ for several values of $r$. For a given index $a$, the limit of this function as $x \to \infty$ is $\lambda_a$. As in Fig.~\ref{fig:appendix_ratios}, each panel corresponds to a fixed value of $r$, and the different colors represent different indices $a$. 

Notice that for each $a$, the limiting value of the curve being plotted is independent of $r$. The resulting $\lambda_a$ are listed in Table~\ref{tab:lambda}.
\begin{table}[h]
\setlength\tabcolsep{5pt}
\begin{tabular}{|lllllll|} 
    \hline 
    $\lambda_{1\rho}$ & $\lambda_{2\rho}$ & $\lambda_{x\rho}$ & $\lambda_{1\sigma}$ &  $\lambda_{x\sigma}$ & $\lambda_{S}$ & $\lambda_{T}$ \\
\hline
$1/2$ & $1/2$  &  $1/2$  & $1/2$  &  $1$  & $1/4$ & $5/4$  \\
    \hline 
\end{tabular}
\caption{\label{tab:lambda}
Exponents for the subdominant couplings. }
\end{table}

\section{Computing $M(g)$
\label{app:M(g)}}
Here we consider the $SU(2)$-symmetric sine-Gordon Hamiltonian which governs $\phi_{2\sigma}$ along the C2S1 ray:
\begin{equation}
\mathcal{H}_{SG} = \mathcal{H}_0 + 2\pi v_2 g \mathcal{H}_{2\sigma}
\end{equation}
in which $\mathcal{H}_0$ is the piece of (\ref{eq:H0}) depending on the $(2\sigma)$ fields, and $\mathcal{H}_{2\sigma}$ is given by (\ref{eq:spin_interaction}). We will compute $M(g) \equiv \ev{\cos({\sqrt{2\pi}\phi_{2\sigma}})}$ for $g \ll 1$. To do so, we add a probe field $h$ to the Hamiltonian:
\begin{equation}
    \mathcal{H}_{\text{SG}} \to \mathcal{H}_0 + 2\pi v_2 g \mathcal{H}_{2\sigma} -h \cos(\sqrt{2\pi}\phi_{2\sigma}).
\end{equation}

To evaluate $M(g)$, we will need the RG equations in the presence of arbitrarily small $h$. In this limit, the beta function for $g$ is unaffected. Thus 
\begin{equation}
    \dv{g}{\ell} = g^2 + a_2g^3 + \ldots \label{eq:g_scaling}
\end{equation}
The two-loop coefficient $a_2$ is known to be $-1/2$ ~\cite{Zamolodchikov_1995} but is left arbitrary for now; we will see that it enters the final result only implicitly through the gap $\Delta_{2\sigma}(g)$.

As for the beta function for $h$, its first order term follows from its scaling dimensions $3/2$. All higher order terms also contain at least one power of $h$; in the limit of infinitesimal $h$ only the terms with one such power matter. The general structure is therefore
\begin{equation}
    \dv{h}{\ell} = (3/2 + b_1 g + b_2g^2+ \ldots )h. \label{eq:h_scaling}
\end{equation}
The higher order terms can be interpreted as a running correction to the scaling dimension of $h$. We postpone for now the evaluation of $b_1$, on which our result will explicitly depend.

It is straightforward to show from the RG transformation properties of the free energy of the corresponding 2D statistical mechanics problem that the ``spontaneous magnetization'' is
\begin{equation}
    M(g) \sim e^{-2 \ell(g)} \underset{h\to 0}\lim \frac{h_{\text{end}}(g,h)}{h}, 
\end{equation}
where $\ell(g)$ denotes the amount of RG ``time'' required to scale from a given $g$ up to some fixed $g_{\text{end}}$, and $h_{\text{end}}(g,h)$ is the value of the probe field at $g_{\text{end}}$ given that it starts off equal to $h$. To evaluate the limit, we divide Eq.~(\ref{eq:h_scaling}) by $h$ and  Eq.~(\ref{eq:g_scaling}) to find
\begin{equation}
    \dv{\log(h)}{g} = \pqty{(3/2) \dv{\ell}{g} + \frac{c_1}{g} +\ldots}
\end{equation}
This integrates to
\begin{align}
    \frac{h_{\text{end}}}{h} &= \exp((3/2) \ell(g) -c_1 \log(g) + \ldots) \\
    &\sim g^{-b_1}e^{(3/2)\ell(g)}
\end{align}
and therefore
\begin{equation}
    M(g) \sim g^{-b_1}e^{-(1/2)  \ell(g)}.
\end{equation}
However, we also know that $\Delta_{2\sigma}(g) \sim e^{-\ell(g)}\Lambda$. So
\begin{equation}
    M(g) \sim g^{-b_1} (\Delta_{2\sigma}(g)/\Lambda)^{1/2}.
\end{equation}
Notice that for $g \ll $ this relationship holds both in the one-loop approximation, and in higher order approximations where $\Delta_{2\sigma}(g)$ acquires an algebraic prefactor.

It remains to evaluate $b_1$. This coefficient is fixed by the operator product expansion (OPE) of $\mathcal{H}_{2\sigma}$ with $\cos(\sqrt{2\pi}\phi_{2\sigma})$. Letting $C$ be the OPE coefficient which appears as
\begin{equation}
    \mathcal{H}_{2\sigma} \times \cos(\sqrt{2\pi}\phi_{2\sigma}) = C \cos(\sqrt{2\pi}\phi_{2\sigma}) + \ldots ,
\end{equation}
we have $b_1 = \alpha C$ for some constant $\alpha$. It is straightforward to evaluate $C$ and $\alpha$. However, we can also read off $b_1$ from the known RG equations for the two-leg ladder, Eq.~(\ref{eq:RG}). This works as follows. The beta function for $g_S$ contains a term $-(3/4)g_{2\sigma} g_S = (3/4)g g_S$. The coefficient in this term is fixed by the OPE coefficient $\widetilde{C}$, where
\begin{equation}
    \mathcal{H}_{2\sigma} \times \mathcal{H}_S =  \widetilde{C} \mathcal{H}_S + \ldots.
\end{equation}
However, $\widetilde{C} = C$, because $\mathcal{H}_{2\sigma}$ depends only on the $(2\sigma)$ fields, but $\cos(\sqrt{2\pi}\phi_{2\sigma})$ and $\mathcal{H}_S$ differ by a factor which is independent of the $(2\sigma)$ fields. Thus, $b_1 = 3/4$, and 
\begin{equation}
    M(g) \sim g^{-3/4} (\Delta_{2\sigma}(g)/\Lambda)^{1/2}.
\end{equation}
As an aside, it is clear that the term $-(3/4)g_{2\sigma} g_S$ in the beta function for $g_S$ is the one responsible for the fact that $g_S \sim g^{3/4}$ along the C2S1 ray. Thus it is not surprising that the explicit $g$-dependence in Eq.~(\ref{eq:gaps_before_evaluating}) cancels.

\section{Remaining sinusoidal interactions \label{app:remaining_interactions}}
The remaining sinusoidal interactions are
\begin{align}
    \mathcal{H}_{x\sigma} &\propto \nonumber \\ &\cos(\sqrt{2\pi}(\phi_{1\sigma}+\phi_{2\sigma}))\cos(\sqrt{2\pi}(\theta_{1\sigma}-\theta_{2\sigma})) + \ldots
\end{align}
\begin{align}
    \mathcal{H}_{x\sigma} &\propto  \cos(\sqrt{4\pi}\theta_{-\rho}) \Big[\sin(\sqrt{2\pi}\phi_1)\sin(\sqrt{2\pi}\phi_2)   \nonumber \\
    &+ \cos(\sqrt{2\pi}(\theta_{1\sigma}-\theta_{2\sigma}))\Big] + \ldots
\end{align}
where $\ldots$ signifies the gradient piece of the interaction. Along the C2S1 ray, the dominant part of the Hamiltonian -- namely $\mathcal{H} + 2\pi v_2 g\mathcal{H}_{2\sigma}$ -- pins $\phi_{2\sigma}$ about $n \sqrt{\pi/2}$, for integer $n$.  Because a pinned $\phi_{2\sigma}$ means $\theta_{2\sigma}$ is wildly fluctuating, and also due to the particular value at which $\phi_{2\sigma}$ is pinned, the terms above vanish when functions of the $(2\sigma)$ field are replaced by their expectation value.

\section{Numerical details \label{app:numerical_details}}
We study the two-leg Hubbard ladder using DMRG with SU(2) spin rotational symmetry. In most of the simulations we keep up to 8000 SU(2) states to reach the typical truncation error $\epsilon\sim 10^{-7}$. To obtain more accurate long range properties of the ground state, we apply finite truncation error extrapolation to all physical quantities we measure.
This works as follows. We compute a given physical quantity $\phi$ as a function of truncation error $\epsilon$ by keeping for each $\epsilon$ the corresponding number of block-states. We then fit $\phi(\epsilon)$ to a second order polynomial $a_0 +a_1\epsilon + a_2 \epsilon^2$, and report the fitted value of $a_0$ as the zero truncation error limit $\phi(0)$. In Fig.~\ref{fig:DMRG} and Fig.~\ref{fig:n(k)}, this procedure is repeated for each individual data point.
In practice, we have used the quantities measured at $m=4000\sim 8000$ kept $SU(2)$ states to extract the $\epsilon\rightarrow 0$ results. For the longest $L_x=288$ system, we also include the quantities measured with $m = 10000$ SU(2) states.

\begin{figure}
    \centering
    \includegraphics[width=0.6\linewidth ]{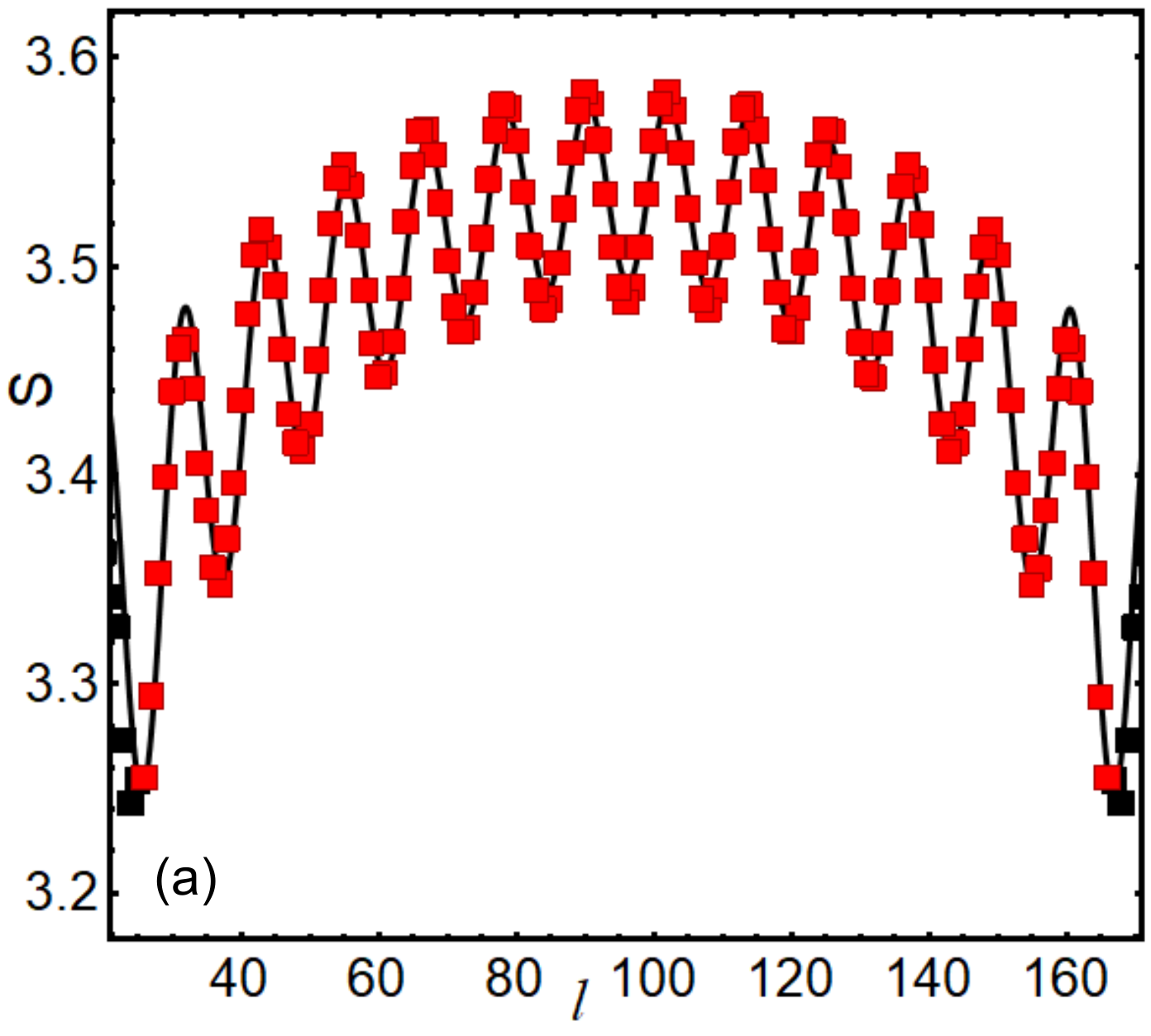}
    \caption{The entanglement entropy $S(l_x)$ for the two-leg ladder at $U=4$, $t_{\perp}=1$, $\delta = 1/12$ and system size $L_x = 192$. The solid line is fit according to Eq.~(\ref{Eq-EE}).
    \label{fig:ee}}
\end{figure}

To determine the number of the gapless mode of the two-leg Hubbard model, we calculate the von Neumann entropy $S(l_x)=-\trace{\rho_{l_x} \ln \rho_{l_x}}$, where $\rho_{l_x}$ is the reduced density matrix of a subsystem with length $l_x$. For critical systems in 1+1 dimensions described by conformal field theory, it has been established \cite{Calabrese2004,Fagotti2011} that for an open system with length $L_x$,
\begin{eqnarray}
S(l_x)&=&\frac{c}{6} \ln \left[\frac{4(L_x+1)}{\pi} \sin \frac{\pi(2l_x+1)}{2(L_x+1)}
\right] \nonumber\\
&+& \frac{A\sin[q(2l_x+1)]}{\frac{4(L_x+1)}{\pi} \sin \frac{\pi(2l_x+1)}{2(L_x+1)}
}+ B. \label{Eq-EE}
\end{eqnarray}
Here $c$ is the central charge, 
and only the leading logarithmic term is universal.  To facilitate the fit to the data, we keep as well non-univesal subleading terms of the form expected of a single gapless mode with $K$ near 1, i.e. for a single gapless Dirac fermion.  In our fits, $c$, $q$, $A$, and $B$ are treated as adjustable parameters, although, as expected on  theoretical grounds, 
we find that $q\to n \pi/2$ as $L_x \to \infty$, where $n$ is the density per site. See Fig.~\ref{fig:ee} for an example fit.

\begin{figure}
    \centering
    \includegraphics[width=7.5 cm]{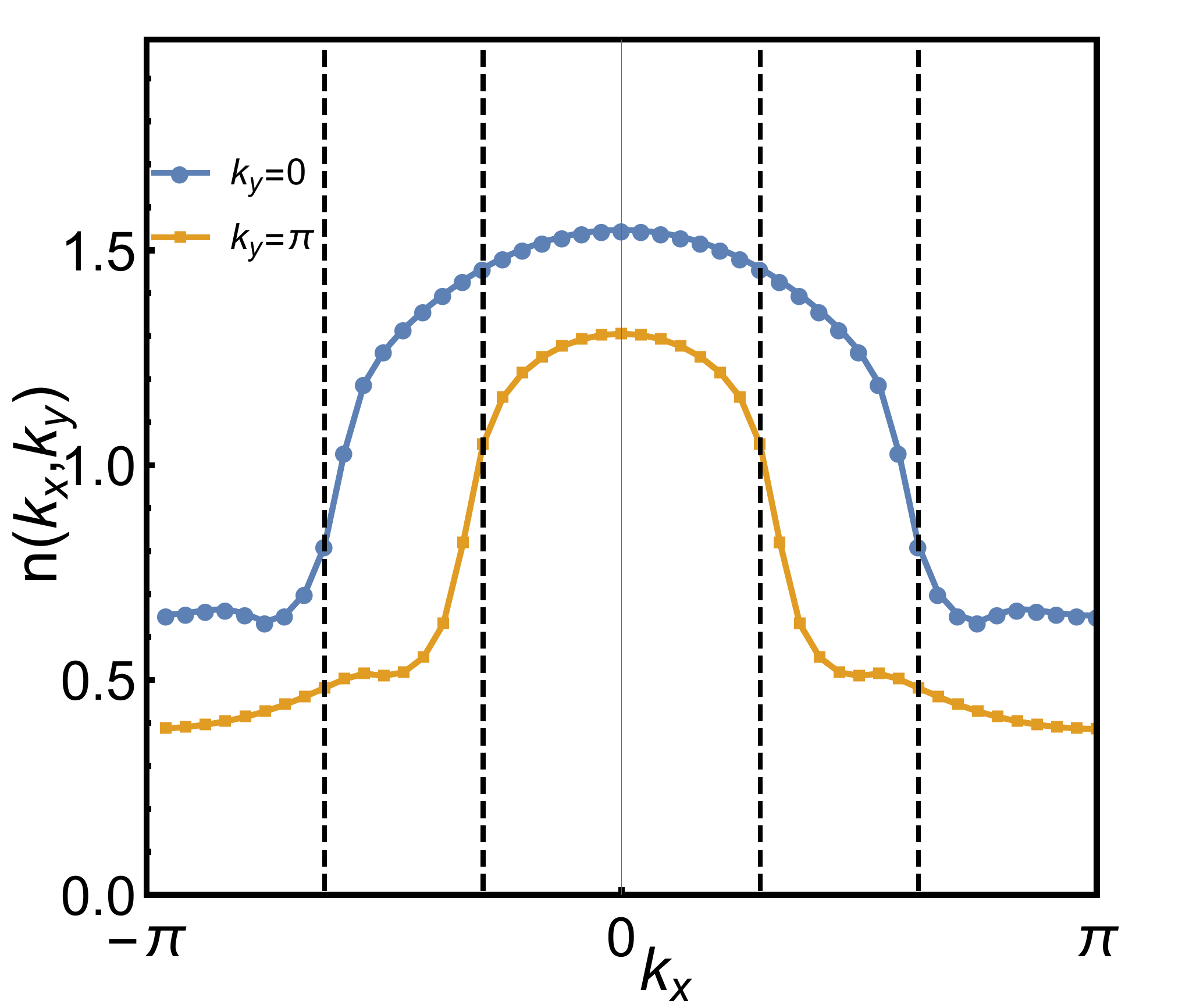}
    \caption{Electron density $n(\bf{k})$ of $L_x=48$ ladder with $U=12$ and doping $1/12$. Dashed lines indicate the Fermi points of non-interacting bands.}
    \label{Afig:nkU12}
\end{figure}

In Fig.~(\ref{Afig:nkU12}), we calculate the $k$-space single-particle occupancy $n(\bf{k})$ of the $L_x=48$ ladder with $U=12$ at doping $\delta=1/12$. Comparing with the sharp drops of $U=4$ case at non-interacting Fermi momenta, the features shown in Fig.~(\ref{Afig:nkU12}) are rounded and their centers are away from the non-interacting momenta.

\bibliography{bibliography}

\end{document}